\documentclass[sigconf, nonacm]{acmart}
\usepackage{footmisc} 
\usepackage{amsthm}
\usepackage{bbm}
\usepackage{subfigure}
\usepackage{url}
\usepackage{booktabs}
\usepackage{enumitem}
\usepackage{booktabs}
\usepackage{caption}
\usepackage{makecell}
\usepackage{multirow}
\usepackage{tcolorbox}
\usepackage{pifont}
\usepackage{xspace}

\newfloat{figtab}{htb}{fgtb}
\makeatletter
  \newcommand\figcaption{\def\@captype{figure}\caption}
  \newcommand\tabcaption{\def\@captype{table}\caption}
\makeatother

\usepackage{enumitem}
\usepackage{pifont}
\usepackage[ruled, vlined, linesnumbered]{algorithm2e}
\usepackage{amsmath}
\usepackage{booktabs}
\usepackage{multirow}

\usepackage{adjustbox}
\usepackage{graphicx}
\usepackage{float}
\usepackage{xcolor}
\SetKwInOut{Parameter}{Parameters}
\usepackage{hyperref}
\def\Snospace~{Section {}}

\usepackage{xspace}

\usepackage{amsmath}
\usepackage{bm}
\usepackage{enumitem}
\usepackage{multirow}
\usepackage{booktabs}
\usepackage{pifont}
\newcommand{\cmark}{\ding{51}}%
\newcommand{\xmark}{\ding{55}}%
\usepackage{lipsum}
\usepackage{bbm}
\usepackage{pifont}
\usepackage{stfloats}

\usepackage[aboveskip=1pt,belowskip=0pt]{caption}

\newcommand{\blue}[1]{\textcolor{blue}{#1}}

\newcommand\vldbdoi{XX.XX/XXX.XX}
\newcommand\vldbpages{XXX-XXX}
\newcommand\vldbvolume{14}
\newcommand\vldbissue{1}
\newcommand\vldbyear{2020}

\newcommand\vldbtitle{\shorttitle} 
\newcommand\vldbavailabilityurl{https://github.com/PKU-DAIR/KnobsTuningEA}
\newcommand\vldbpagestyle{plain}

\begin{document}

\setlength{\textfloatsep}{11.3pt}
\setlength{\floatsep}{5pt}
\title{Facilitating Database  Tuning with Hyper-Parameter Optimization: A Comprehensive Experimental Evaluation }

\subtitle{[Experiment, Analysis \& Benchmark]}


\author{Xinyi Zhang$^{ \dagger\ddagger\ast}$, Zhuo Chang$^{\dagger\ddagger\ast}$, Yang Li$^\dagger$, Hong Wu$^\ddagger$, Jian Tan$^\ddagger$, Feifei Li$^\ddagger$, Bin Cui$^\dagger$}
\affiliation{
$^\dagger$EECS, Peking University~~~~~$^\ddagger$Alibaba Group
}
\affiliation{
$^\dagger$\{zhang\_xinyi, z.chang, liyang.cs, bin.cui\}@pku.edu.cn~~~~~ $^\ddagger$\{hong.wu, j.tan, lifeifei\}@alibaba-inc.com
}


\begin{abstract}

Recently, using automatic configuration tuning to improve the performance of modern database management systems (DBMSs) has attracted increasing interest from the database community. 
This is embodied with a number of systems featuring advanced tuning capabilities being developed.
However, it remains a challenge to select the best solution for database configuration tuning, considering the large body of algorithm choices.
In addition, beyond the applications on database systems, we could find more potential algorithms designed for configuration tuning.
To this end, this paper provides a comprehensive evaluation of configuration tuning techniques from a broader perspective, hoping to better benefit the database community. 
In particular, we summarize three key modules of database configuration tuning systems and conduct extensive ablation studies using various challenging cases.  
Our evaluation demonstrates that the hyper-parameter optimization algorithms can be borrowed to further enhance the database configuration tuning. 
Moreover, we identify the best algorithm choices for different modules.
Beyond the comprehensive evaluations, we offer an efficient and unified database configuration tuning benchmark via surrogates that reduces the evaluation cost to a minimum, allowing for extensive runs and analysis of new techniques.

\end{abstract}


\setcounter{table}{0}
\setcounter{figure}{0}
\setcounter{section}{0}

\settopmatter{printfolios=true}
\maketitle

\pagestyle{\vldbpagestyle}
\begingroup\small\noindent\raggedright\textbf{PVLDB Reference Format:}\\
Xinyi Zhang, Zhuo Chang, Yang Li, Hong Wu, Jian Tan, Feifei Li, Bin Cui. 
\vldbtitle. PVLDB, \vldbvolume(\vldbissue): \vldbpages, \vldbyear.\\
\href{https://doi.org/\vldbdoi}{doi:\vldbdoi}
\endgroup
\begingroup

\renewcommand\thefootnote{}\footnote{\noindent
$^\ast$Xinyi Zhang and Zhuo Chang contribute equally to this paper. \\
\noindent This work is licensed under the Creative Commons BY-NC-ND 4.0 International License. Visit \url{https://creativecommons.org/licenses/by-nc-nd/4.0/} to view a copy of this license. For any use beyond those covered by this license, obtain permission by emailing \href{mailto:info@vldb.org}{info@vldb.org}. Copyright is held by the owner/author(s). Publication rights licensed to the VLDB Endowment. \\
\raggedright Proceedings of the VLDB Endowment, Vol. \vldbvolume, No. \vldbissue\ %
ISSN 2150-8097. \\
\href{https://doi.org/\vldbdoi}{doi:\vldbdoi} \\
}\addtocounter{footnote}{-1}\endgroup

\ifdefempty{\vldbavailabilityurl}{}{
\begingroup\small\noindent\raggedright\textbf{PVLDB Artifact Availability:}\\
The source code, data, and/or other artifacts have been made available at \url{\vldbavailabilityurl}.
\endgroup
}

\section{Introduction}

\begin{figure}[t]
    \centering
    \includegraphics[width=1\linewidth]{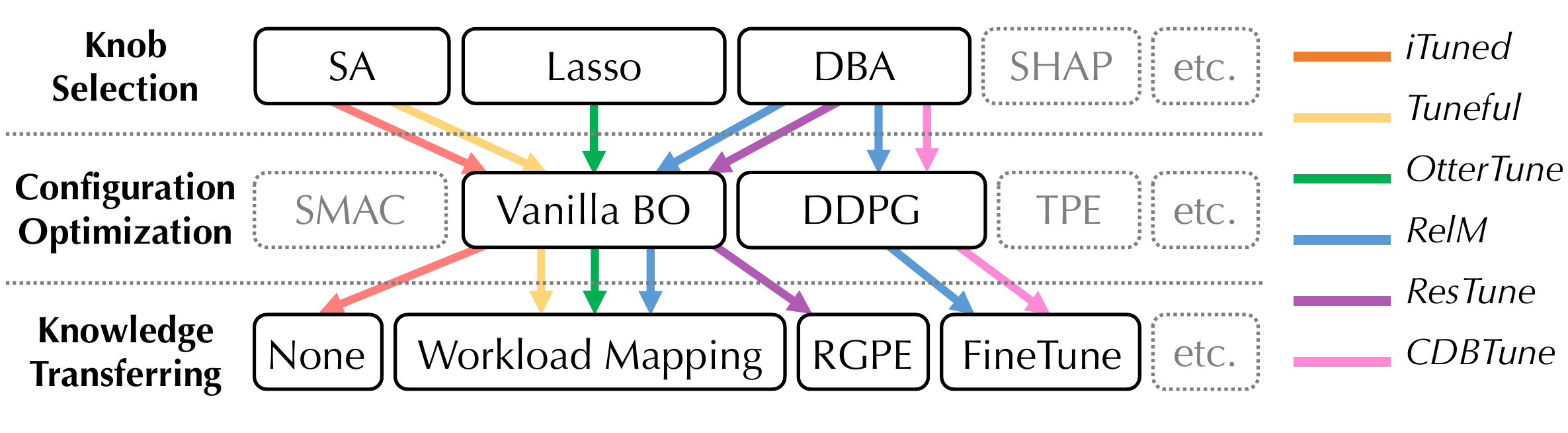}
    \caption{Algorithm Choices of Configuration Tuning Systems: 
    Black boxes denote the algorithms adopted by existing database tuning systems (indicated by colored paths), and grey boxes denote the algorithms in the HPO field. 
    }
    \label{fig:intro}
\end{figure}

Modern database management systems (DBMSs) have hundreds of configuration knobs that determine their runtime behaviors~\cite{DBLP:conf/vldb/ChaudhuriN07}. 
Setting the appropriate values for these configuration knobs is crucial to pursue the high throughput and low latency of a DBMS.
Given a target workload, configuration tuning aims to find configurations that optimize the database performance. 
This problem is proven to be NP-hard~\cite{DBLP:conf/sigmetrics/SullivanSP04}.
To find a promising configuration for the target workload,
database administrators (DBAs) put significant effort into tuning the configurations. Unfortunately, manual tuning struggles to handle different workloads and hardware environments, especially in the cloud environment~\cite{DBLP:journals/pvldb/Pavlo21}. Therefore, automatic configuration tuning attracts intensive interests in both academia and industry~\cite{DBLP:conf/vldb/AgrawalCKMNS04,DBLP:conf/vldb/WeikumMHZ02,DBLP:conf/vldb/StormGLDS06,DBLP:conf/vldb/ShashaB02,DBLP:conf/vldb/ChaudhuriW06,DBLP:conf/vldb/ShashaR92,DBLP:journals/dpd/KossmannS20,DBLP:conf/sigmod/MaAHMPG18}.

Recently, there has been an active research area on automatically tuning database configurations using Machine Learning (ML) techniques~\cite{DBLP:journals/pvldb/DuanTB09, DBLP:conf/sigmod/AkenPGZ17, DBLP:conf/kdd/FekryCPRH20,DBLP:conf/sigmod/ZhangLZLXCXWCLR19,DBLP:journals/pvldb/LiZLG19,DBLP:conf/sigmod/KunjirB20,DBLP:conf/sigmod/ZhangWCJT0Z021,DBLP:conf/sigmod/MaAHMPG18}. 
We summarize three key modules in the existing tuning systems:  \textit{knob selection} that prunes the configuration space, \textit{configuration optimization} that samples promising configurations over the pruned space, and \textit{knowledge transfer} that further speeds up the tuning process via historical data.
Based on the techniques used in \textit{configuration optimization} module, these systems can be categorized into two major types: Bayesian Optimization (BO) based systems~\cite{DBLP:journals/pvldb/DuanTB09,DBLP:conf/sigmod/AkenPGZ17,DBLP:conf/kdd/FekryCPRH20,DBLP:conf/sigmod/ZhangWCJT0Z021} and Reinforcement Learning (RL) based~\cite{DBLP:conf/sigmod/ZhangLZLXCXWCLR19, DBLP:journals/pvldb/LiZLG19} systems. 
Owing to these efforts, modern database systems are well-equipped with powerful algorithms for configuration tuning.
Examples includes  Lasso algorithm in OtterTune~\cite{DBLP:conf/sigmod/AkenPGZ17}, Sensitivity Analysis (SA) in Tuneful ~\cite{DBLP:conf/kdd/FekryCPRH20} for automatic \textit{knob selection}; Bayesian Optimization (e.g., iTuned~\cite{DBLP:journals/pvldb/DuanTB09}, ResTune~\cite{DBLP:conf/sigmod/ZhangWCJT0Z021}), Reinforcement Learning (e.g., CDBTune~\cite{DBLP:conf/sigmod/ZhangLZLXCXWCLR19}, Qtune~\cite{DBLP:journals/pvldb/LiZLG19}) for \textit{configuration optimization}; and  workload mapping in OtterTune, RGPE~\cite{feurer2018scalable}  in ResTune~\cite{DBLP:conf/sigmod/ZhangWCJT0Z021} for \textit{knowledge transfer}.
The various algorithms in each module of the configuration tuning systems enrich the solutions for database optimization and demonstrate superior performance and efficiency compared to manual tuning.

\subsection{Motivation}
Whilst a large body of methods have been proposed, a comprehensive evaluation is still missing.
Existing evaluations~\cite{DBLP:journals/pvldb/DuanTB09,DBLP:conf/sigmod/AkenPGZ17,DBLP:conf/sigmod/ZhangLZLXCXWCLR19,DBLP:journals/pvldb/LiZLG19,DBLP:conf/sigmod/ZhangWCJT0Z021} compare tuning systems from a macro perspective, lacking analysis of algorithm components in various scenarios. 
This motivates us to conduct a comprehensive comparative analysis and experimental evaluation of database tuning approaches from a micro perspective. We now discuss the issues of the existing work.

\vspace{0.3em}
\noindent\textbf{I1: Missing comparative evaluations of intra-algorithms in different modules.}
Emerging database tuning systems are characterized by new intra-algorithms (i.e., algorithms in each module) such as
OtterTune with Lasso-based knob selection and workload mapping, CDBTune with DDPG optimizer, ResTune with RGPE transfer framework.
Figure \ref{fig:intro} plots the three key modules and the corresponding intra-algorithms we extracted from the designs of these systems.
When designing a database tuning system, we can construct many possible ``paths'' across the intra-algorithm choices among the three modules, even not limited to existing designs.
For example, each \textit{knob selection} algorithm determines a unique configuration space and can be ``linked'' to any of the \textit{configuration optimization} algorithms.
Given so many possible combinations, it remains unclear to identify the best ``path'' for database configuration tuning in practice.
Existing researches focus on the evaluation of the entire tuning systems~\cite{DBLP:conf/sigmod/AkenPGZ17, DBLP:conf/kdd/FekryCPRH20, DBLP:conf/sigmod/ZhangLZLXCXWCLR19} or limited intra-algorithms in some of the modules~\cite{DBLP:journals/pvldb/AkenYBFZBP21}, failing to reveal which intra-algorithm contributes to the overall success.
For example, the choice of intra-algorithms in \textit{knob selection} module is often overlooked, yet important, since different algorithms can lead to distinct configuration spaces, affecting later optimization.
To this end, it is essential to conduct a thorough evaluation with system breakdown and fine-grained intra-algorithm comparisons.

\vspace{0.3em}
\noindent\textbf{I2: Absence of analysis for high-dimensional and heterogeneous scenarios.}
There are two challenging scenarios for configuration optimization: \textit{high dimensionality} and \textit{heterogeneity} of configuration space.
DBMS has hundreds of configuration knobs that could be continuous (e.g., \textit{innodb\_buffer\_pool\_size} and \textit{tmp\_table\_size}) or categorical (e.g., \textit{innodb\_stats\_method} and \textit{innodb\_flush\_neighbors}). 
We refer to the scenario with knobs of various types as \textit{heterogeneity}.
And, the categorical knobs vary fundamentally from the continuous ones in differentiability and continuity.
For instance, existing BO-based methods tend to yield the best results for low-dimensional and continuous spaces.
However, as for the high-dimensional space, they suffer from the over-exploration issue~\cite{DBLP:journals/pieee/ShahriariSWAF16}.
In addition, vanilla BO methods assume a natural ordering of input value~\cite{DBLP:conf/icml/WanNHRLO21}, thus struggling to model  the heterogeneous space.
Consequently, when analyzing the optimizers, it is essential to compare their performance under the two scenarios.

\vspace{0.3em}
\noindent\textbf{I3: Limited solution comparison without a broader view beyond database community.}
The database configuration tuning is formulated as a
black-box optimization problem over configuration space.
Thanks to the efforts of recent studies, today’s database practitioners are well-equipped with many techniques of configuration tuning.
However, when we look from a broader view, we can find plenty of toolkits and algorithms designed for the black-box optimization, especially  hyper-parameter optimization (HPO) approaches~\cite{DBLP:journals/ijon/YangS20,DBLP:conf/avss/MauriceML17,DBLP:conf/clei/BarscePM17,DBLP:conf/cec/0025LB19,DBLP:conf/iwbf/Gonzalez-Cuautle19,mfeshb21li}.
HPO aims to find the optimal hyper-parameter configurations of a machine learning algorithm as rapidly as possible to minimize the corresponding loss function.
Database configuration tuning shares a similar spirit with HPO since they both optimize a black-box objective function with expensive function evaluations.
We notice that recent advances in HPO field have shown promising improvement in high-dimensional and heterogeneous configuration spaces~\cite{DBLP:conf/lion/HutterHL11, DBLP:conf/nips/ErikssonPGTP19,DBLP:conf/aaai/KimLLKL20,DBLP:conf/aistats/WangSJF14,DBLP:conf/aistats/WangGKJ18}.
Despite their success, such opportunities to further facilitate database configuration tuning have not been investigated in the literature.

\vspace{0.3em}
\noindent\textbf{I4: Lack of cheap-to-evaluate and unified database tuning benchmarks.}
Evaluating new algorithms in database tuning systems can be costly, time-consuming, and hard to interpret. 
It requires DBMS copies, computing resources, and the infrastructure to replay workloads and the tools to collect performance metrics.
It takes dozens of hours to conduct a single run of optimization with ceaseless computing resources.
Moreover, obtaining a sound evaluation of more baselines takes several-fold times longer with expensive computing costs.
In addition, the database performance can fluctuate across instances of the same type even though the configuration and workload are the same~\cite{DBLP:journals/pvldb/AkenYBFZBP21}.
Therefore, we cannot run the optimizers in parallel to reduce the evaluation time, making the evaluation more troublesome.
These problems pose a considerable barrier to the solid evaluation of database tuning systems.
An efficient and unified tuning benchmark is needed. 

\begin{table*}[t]
\caption{Taxonomy and Brief Description of Existing Database Configuration Tuning Systems.}\label{tab:system}
\scalebox{0.95}[0.95]{

\begin{tabular}{|l|c|c|c|}
\hline
Category                  & Configuration Tuning System                 & Application                             & Design Highlight                             \\ \hline
\multirow{6}{*}{BO-based} & iTuned \cite{DBLP:journals/pvldb/DuanTB09}  & Performance tuning for DBMS             & First adopting BO                            \\ \cline{2-4} 
                          & OtterTune \cite{DBLP:conf/sigmod/AkenPGZ17} & Performance tuning for DBMS             & Incremental knob selection, workload mapping \\ \cline{2-4} 
 &
  Tuneful \cite{DBLP:conf/kdd/FekryCPRH20} &
  Performance tuning for analytics engines &
  Incremental knob selection via Sensitivity Analysis \\ \cline{2-4} 
 &
  ResTune \cite{DBLP:conf/sigmod/ZhangWCJT0Z021} &
  Resourse-oriented tuning for DBMS &
  Adopting RGPE to transfer historical knowledge \\ \cline{2-4} 
                          & RelM \cite{DBLP:conf/sigmod/KunjirB20}      & Memory allocation for analytics engines & Combining white-box knowledge                \\ \cline{2-4} 
 &
  CGPTuner \cite{DBLP:journals/pvldb/CeredaVCD21} &
  Performance tuning for IT systems &
  Adopting Contextual BO to adapt to workload variation \\ \hline

\multirow{2}{*}{RL-based} &
  CDBTune \cite{DBLP:conf/sigmod/ZhangLZLXCXWCLR19} &
  Performance tuning for DBMS &
  First adopting DDPG \\ \cline{2-4} 
                          & QTune \cite{DBLP:journals/pvldb/LiZLG19}    & Performance tuning for DBMS             & Supporting three  tuning granularities       \\ \hline
\end{tabular}
}
\end{table*}

\subsection{Our Contributions}
Driven by the aforementioned issues, we provide comprehensive analysis and experimental evaluation of database configuration tuning approaches.
Our contributions are summarized as follows:

\vspace{0.3em}
\noindent\textbf{C1. We present a unified pipeline with three key modules and evaluate the fine-grained intra-algorithms.}
For \textbf{I1}, we break down all the database tuning systems into three key modules and evaluate the corresponding intra-algorithms.
We evaluate and analyze each module's intra-algorithms to answer the motivating questions: 
\textit{(1) How to determine tuning knobs? (2) Which optimizer is the winner? (3) Can we transfer knowledge to speed up the target tuning task?}
With our in-depth analysis, we identify design trade-offs and potential research opportunities. 
We discuss the best ``paths'' across the three modules in various scenarios and present all-way guidance for configuration tuning in practice. 

\vspace{0.3em}
\noindent\textbf{C2. We construct extensive scenarios to benchmark multiple optimizers.} 
For \textbf{I2}, we construct specific evaluations and carry out comparative analysis for the two challenges.
For \textit{high dimensionality}, we conduct evaluations of the optimizers over configuration spaces of three sizes: small, medium, and large.
We specifically analyze the different performances when optimizing over small/medium and large configuration spaces. 
For \textit{heterogeneity}, we construct a comparison experiment with continuous and heterogeneous configuration spaces to validate the optimizers' support for \textit{heterogeneity}. 
Such evaluation setting assists us in better understanding the pros and cons of existing database optimizers.

\vspace{0.3em}
\noindent\textbf{C3. Think out of the box: we apply and evaluate advanced HPO techniques in database tuning problems.}
For \textbf{I3}, we survey existing solutions for black-box optimization.
We find that many recent approaches in the HPO field could be borrowed to alleviate the challenges in database tuning. 
Therefore, besides the configuration tuning techniques proposed in the database community, we also evaluate other advanced approaches in the HPO field under the same database tuning setting.
Through evaluations, we have that the best knob selection approach achieves 38.02\% average performance improvement, and the best optimization approach achieves 21.17\% average performance improvement as well as significant speedup compared with existing methods.
We demonstrate that database practitioners can borrow strength from the HPO approaches in an out-of-the-box manner.
 
\vspace{0.3em}
\noindent\textbf{C4. We define an efficient database configuration tuning benchmark via surrogates.}
We summarize the difficulties discussed in \textbf{I4} lie in the cost and fluctuation of performance evaluations under configurations. 
We propose benchmarking database configuration tuning via a regression surrogate that approximates future evaluations through cheap and stable model predictions~\cite{DBLP:conf/aaai/EggenspergerHHL15}.
Specifically, we train regression models on \textit{(configuration, performance)} pairs collected in an expensive offline manner and cheaply evaluate future configurations using the model’s performance predictions instead of replying workloads.
The tuning benchmark offers optimization scenarios that share the same configuration spaces and feature similar response surfaces with real-world configuration tuning problems.
Significantly, it reduces the evaluation cost to a minimum and achieves 150\textasciitilde 311 $\times$ overall speedup.  
The benchmark is publicly available to facilitate future research.
With the help of the benchmark, researchers can skip time-consuming workload replay, conduct algorithm analysis and comparison efficiently, and test new algorithms with few costs.

  \begin{figure}
\centering
    \includegraphics[keepaspectratio=true,scale=0.38]{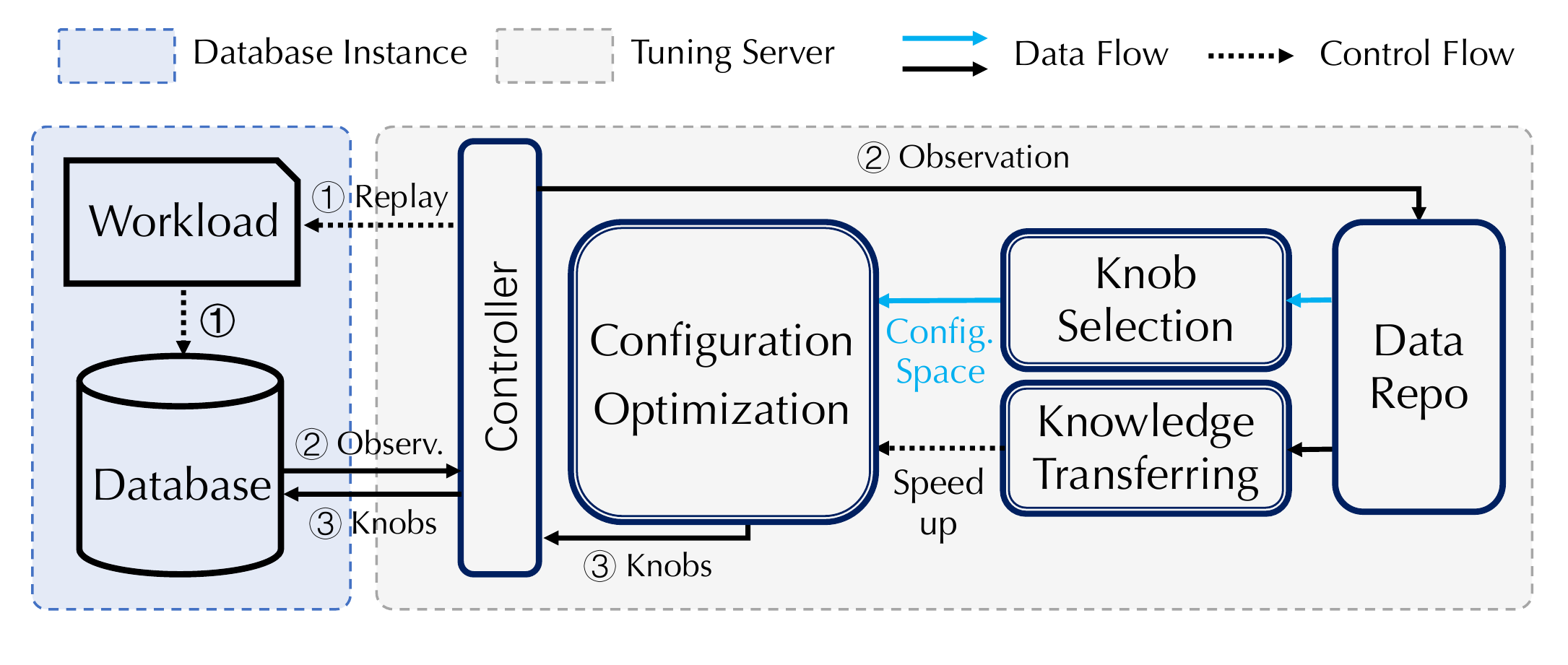}
\caption{ Architecture  of  Configuration Tuning System.}
\vspace{-0.5em}
\label{fig:sys}

\end{figure}

The rest of the paper is organized as follows: 
Section \ref{sec2} introduces the preliminaries of this paper, demonstrating the taxonomy and workflow of existing database tuning systems as well as the background knowledge for HPO.
Next, we present a survey on algorithms for the three modules in Section \ref{sec3} and describe the general  setup in Section \ref{sec4}.
Then, the experimental  analysis are presented in Sections \ref{sec5}-\ref{sec:exp-transfer}, followed by the the database tuning benchmark in Section \ref{sec8}.
We summarize the lessons learned and  research opportunities in Section \ref{sec9}. 
Finally, we conclude the paper.

\vspace{-1em}

\section{Preliminaries}\label{sec2}
We first formalize the database configuration tuning problem.
Then we review the architecture and workflow of database configuration tuning systems and present the taxonomy and description about the existing systems. 
Finally, we introduce the background of hyper-parameter optimization.

\subsection{Problem Statement}\label{sec:problem}
Consider a database system having configuration knobs  $\theta_1, \dots,\theta_m$, we define its configuration space $\boldsymbol{\Theta}=\Theta_1 \times \dots \times \Theta_m $, where $\Theta_1,\dots,\Theta_m$ denote their respective domains. 
A configuration knob $\theta$ can be continuous or categorical.
We denote the database performance metric as $f$ which can be any chosen metric to be optimized, such as \textit{throughput}, \textit{99\%th percentile latency}, etc.
Given a workload and a specific configuration $\boldsymbol{\theta}$, the corresponding performance $f(\boldsymbol{\theta})$ can be
observed after evaluating it in the database.
We denote the historical observations as $H_t=\{(\boldsymbol{\theta_i},y_i)\}_1^{t}$. 
Assuming that the objective is a maximization
problem, database configuration tuning aims to find a configuration $\boldsymbol{\theta}^{*}\in\boldsymbol{\Theta}$, where
\begin{equation}
\boldsymbol{\theta}^{*} =  \mathop{\arg\max}_{\boldsymbol{\theta} \in \boldsymbol{\Theta}} f(\boldsymbol{\theta}).
\end{equation}

\subsection{System Overview and Workflow}
The database configuration tuning systems share a typical architecture as shown in Figure \ref{fig:sys}.
The left part shows the target DBMS and the replayed workload.
The right part represents the tuning server deployed in the backend tuning cluster. 
The tuning server maintains a repository of historical observations from tuning tasks. 
The controller monitors the states of the target DMBS and transfers the data between the DBMS side and the tuning server side. 
It conducts stress testing to the target DBMS through standard workload testing tools (e.g., OLTP-Bench~\cite{DBLP:journals/pvldb/DifallahPCC13}) or workload runner that executes the current user's real workload~\cite{DBLP:conf/nips/SnoekLA12}. 
There are three main modules in the tuning system: (1) \textit{knob selection}, (2) \textit{configuration optimization}, (3) \textit{knowledge transfer}. 
The \textit{knob selection} module prunes the configuration space by identifying important knobs.
The pruned configuration space is passed to the optimizer, which suggests a promising configuration over the pruned space at each iteration.
Furthermore, the \textit{knowledge transfer} module speeds up the current tuning task by borrowing strength from past similar tasks stored in the data repository or fine-tuning the optimizer's model.

The \textit{configuration optimization} module functions iteratively, as shown in the black arrows: 
(1) The controller replays the workload to conduct stress testing.
(2) After the stress testing finishes, the performance statistics are uploaded to the data repository, and
is used for updating the model in the configuration optimizer. 
Optionally, the knowledge transfer module can speed up the optimization with historical observations.
(3) The optimizer generates a promising configuration over the pruned configuration space, and the configuration is applied to the target DBMS through the controller.
The above procedures repeat until the model of the optimizer
converges or stop conditions are reached (e.g., budget limits).

\subsection{Taxonomy}\label{sec2.3}
Table \ref{tab:system} presents a taxonomy of existing configuration tuning systems with ML-based techniques.
We classify the systems into two classes based on the algorithms of their optimizers: (1) BO-based systems and (2) RL-based systems.

\subsubsection{\underline{BO-based}}
The existing BO-based systems adopt BO
to model the relationship between configurations and database performance.
They follow BO framework to search for an optimal DBMS configuration: (1) fitting probabilistic surrogate models and (2) choosing the next configuration to evaluate by maximizing acquisition function.

\noindent\textbf{iTuned~\cite{DBLP:journals/pvldb/DuanTB09}}
first adopts vanilla BO models to search for a well-performing configuration. It uses a stochastic sampling technique -- Latin Hypercube Sampling~\cite{DBLP:conf/wsc/McKay92} for initialization and does not use the observations collected from previous tuning sessions to speed up the target tuning task.

\noindent\textbf{OtterTune~\cite{DBLP:conf/sigmod/AkenPGZ17}} selects the most impactful knobs with Lasso algorithm, and increases the size of configuration space  incrementally.
It proposes to speed up the target tuning by a transfer framework via workload mapping. 
Concretely, it re-uses the historical observations of a prior workload by mapping the target workload based on the measurements of internal metrics in DBMS.

\noindent\textbf{Tuneful~\cite{DBLP:conf/kdd/FekryCPRH20}} proposes a tuning cost amortization model and demonstrates its advantage when tuning the recurrent workloads.
It conducts incremental knob selection via Gini score as an importance measurement to gradually decrease the configuration space.
Similar to OtterTune, it also adopts the workload mapping framework to transfer knowledge across tuning tasks.

\noindent\textbf{ResTune~\cite{DBLP:conf/sigmod/ZhangWCJT0Z021}} defines a resource-oriented tuning problem and adopts a constrained Bayesian optimization solver.
It uses an ensemble framework (i.e., RGPE) to combine workload encoders across tuning tasks to transfer historical knowledge.

\noindent\textbf{RelM~\cite{DBLP:conf/sigmod/KunjirB20}} is designed to configure the memory allocation for distributed analytic engines, e.g., Spark.
It builds a white-box model of memory allocation to speed up Bayesian Optimization.

\noindent\textbf{CGPTuner~\cite{DBLP:journals/pvldb/CeredaVCD21}} proposes to  tune the configuration of an IT system, such as  the Java Virtual Machine (JVM) or the  Operating System (OS)) to unlock the full performance potential of a DBMS.
It adopts Contextual BO~\cite{DBLP:conf/nips/KrauseO11} to adapt  to workload variation, which requires an external workload characterisation module.

\subsubsection{\underline{RL-based}}
RL-based systems adopt Deep Deterministic Policy Gradient (DDPG), which is a policy-based
model-free reinforcement learning agent. 
It treats knob tuning as a trial-and-error procedure to trade-off between exploring unexplored space and exploiting existing knowledge.

\noindent\textbf{CDBTune~\cite{DBLP:conf/sigmod/ZhangLZLXCXWCLR19}} first adopts DDPG algorithm to tune the database.
Its tuning agent inputs internal metrics of the DBMS and outputs proper configuration by modeling the tuning process as a Markov Decision Process (MDP)~\cite{1957A}. 

\noindent\textbf{QTune~\cite{DBLP:journals/pvldb/LiZLG19}} supports three tuning granularities, including workload-level tuning, query-level tuning, and cluster-level tuning.
It embeds workload characteristics to predict the internal metrics for query-level tuning, and clusters the queries based on their ``best'' knob values for cluster-level tuning.
\\ \hspace*{\fill} \\
\noindent\underline{\textbf{Scope Illustration.}}
To make our evaluation focused yet comprehensive, we employ some necessary constraints.
We focus on evaluating the configuration tuning techniques  at the workload level since query-lever or cluster-lever tuning are only available on QTune.
In this paper, we concentrate on performance tuning in DBMS.
As for the systems targeting the other applications (i.e., performance tuning for analytic engines, resource-oriented tuning),
we implement their core techniques (e.g., RGPE transfer framework) and evaluate them in a unified setting.

\subsection{Hyper-Parameter Optimization}
HPO aims to find the  hyper-parameter configurations $\lambda^{*}$ of a given algorithm $A^{i}$ with the best performance on the validation set~\cite{DBLP:conf/mod/GeitleO19}:
\begin{equation}
\small
 \lambda^{*} = \mathop{\arg\min}_{\lambda \in \Lambda_{i}} L(A^{i}, \lambda).
\end{equation}
HPO is known as black-box optimization since nothing is known about the loss function besides its function evaluations.
BO has been successfully applied to solve the HPO problem~\cite{DBLP:conf/iisa/MaluDS21,DBLP:journals/jair/WangHZMF16,DBLP:conf/nips/BelakariaDD19}.
The main idea of BO is to use a probabilistic surrogate model to describe the relationship between a hyper-parameter configuration and its performance (e.g., validation error), and then utilize this surrogate to guide the configuration search~\cite{DBLP:journals/ijns/Seeger04}.
It is common to use Gaussian processes (GP) as surrogates~\cite{DBLP:conf/nips/SnoekLA12,DBLP:journals/jmlr/Martinez-Cantin14} and
SMAC~\cite{DBLP:conf/lion/HutterHL11} and TPE~\cite{DBLP:conf/nips/BergstraBBK11} are two other well-established methods.
\section{Methodologies}\label{sec3}
In this section, we survey the methodologies for the three components, respectively.
We not only cover the techniques used in existing configuration tuning systems for DBMS but also discuss the representative approaches from the HPO field.

\subsection{Knob Selection}\label{sec:men-knob}

\begin{table}[t]
\setlength\tabcolsep{2pt}
\small
\caption{Importance Measurements in Knob Selection.}\label{tab:knobs}
\scalebox{0.9}[0.9]{

\begin{tabular}{|c|c|c|}
\hline
Measure                                              & Category                                                   & Brief Description                                                                                                                                \\ \hline
Lasso~\cite{tibshirani1996regression}                                                    & \begin{tabular}[c]{@{}c@{}}Variance\\ based\end{tabular}   & \begin{tabular}[c]{@{}c@{}}Based on coefficient of linear regression,\\ effective for  when existing irrelevant features.\end{tabular}           \\ \hline
\begin{tabular}[c]{@{}c@{}}Gini\\ score~\cite{DBLP:journals/bioinformatics/NembriniKW18}\end{tabular}     & \begin{tabular}[c]{@{}c@{}}Variance\\ based\end{tabular}   & \begin{tabular}[c]{@{}c@{}}Based on the times a feature is used in tree splits,\\ successful in high-dimensional feature selection.\end{tabular} \\ \hline
fANOVA~\cite{DBLP:conf/icml/HutterHL14}                                                   & \begin{tabular}[c]{@{}c@{}}Variance\\ based\end{tabular}   & \begin{tabular}[c]{@{}c@{}}Decomposing the variance of the target function,\\ commonly used in the HPO field.\end{tabular}                            \\ \hline
\begin{tabular}[c]{@{}c@{}}Ablation\\ analysis\cite{DBLP:conf/aaai/BiedenkappLEHFH17}\end{tabular} & \begin{tabular}[c]{@{}c@{}}Tunability\\ based\end{tabular} & \begin{tabular}[c]{@{}c@{}}Comparing feature difference between configurations,\\ straightforward and intuitive.\end{tabular}                     \\ \hline
SHAP~\cite{DBLP:conf/nips/LundbergL17}                                                     & \begin{tabular}[c]{@{}c@{}}Tunability\\ based\end{tabular} & \begin{tabular}[c]{@{}c@{}}Decomposing the performance change additively,\\ solid theoretical foundation derived from game theory.\end{tabular}    \\ \hline
\end{tabular}}
\end{table}

The \textit{knob selection} module identifies and selects important knobs to tune.
Although the database system has hundreds of knobs, not all of them significantly impact database performance.
Selecting important knobs can prune the configuration space and further accelerate configuration optimization. 

To conduct knob selection, we first need to collect a set of observations under different configurations.   
Given the observations, we adopt an algorithm to rank the knobs in terms of their importance. 
(We refer to the ranking algorithm as ``importance measurement'' to distinguish it from the other techniques.)
Finally, the configuration space is determined by selecting the top-k knobs according to the importance rank.
There are various choices for importance measurements, leading to distinct configuration spaces and affecting later optimization.
They can be classified into two categories as shown in Table \ref{tab:knobs}: \textit{variance-based} measurements and \textit{tunability-based} measurements~\cite{DBLP:journals/corr/abs-2007-07588}.
\textit{Variance-based} measurement selects the knobs that have the largest impact on the database performance.
It has been adopted by existing tuning systems for DBMS. 
\textit{Tunability-based} measurement~\cite{DBLP:journals/jmlr/ProbstBB19} quantifies the tunability of a knob, measuring the performance gain that can be achieved by tuning the knob from its default value.
It has been applied successfully to determine the importance of hyper-parameters of ML algorithms, especially when given a well-performing default configuration~\cite{DBLP:journals/corr/abs-2007-07588}.

\subsubsection{\underline{Variance-based Measurements}}
We introduce three variance-based measurements due to their widely-used applications and representativeness, including  Lasso~\cite{tibshirani1996regression} adopted in OtterTune, Gini score~\cite{DBLP:journals/bioinformatics/NembriniKW18} adopted in Tuneful, and functional ANOVA~\cite{DBLP:conf/icml/HutterHL14}, a state-of-the-art importance measurements from the HPO domain.

\noindent\textbf{Lasso~\cite{tibshirani1996regression}} uses the coefficient of linear regression to assess the importance of knobs.
It penalizes the L1-norm of the coefficients and forces the coefficients of  redundant knobs to be zero. 
The L1-norm penalization makes Lasso effective when there are many irrelevant knobs in the training samples~\cite{efron2004least}.
However, Lasso assumes the linearity of the knob space~\cite{krakovska2019performance} and is not able to  capture the non-linear dependencies from knobs to the performance metric.

\noindent\textbf{Gini Score~\cite{DBLP:journals/bioinformatics/NembriniKW18}} is derived from tree-based models, like random forest model~\cite{DBLP:journals/jcisd/SvetnikLTCSF03}.
The Gini score of each knob is defined as the number of times the given knob is used in a tree split across all the  trees since important knobs discriminate the larger number of samples and are used more frequently in  tree splits.
Gini score has been successfully applied to high-dimensional feature selection ~\cite{DBLP:journals/bmcbi/MenzeKMHBPH09}.

\noindent\textbf{fANOVA~\cite{DBLP:conf/icml/HutterHL14}} (i.e., Functional analysis of variance) measures the importance of knobs by analyzing how much each knob contributes to the variance of the target function across the configuration space.
Based on a regression model (e.g., random forest), functional ANOVA  decomposes the target function into additive components that only depend on subsets of its inputs. 
The importance of each knob is quantified by the fraction of variance it explains.
Functional ANOVA is commonly used in the HPO domian due to its solid theoretical foundation~\cite{DBLP:conf/kdd/RijnH18,DBLP:journals/netmahib/Luo16,DBLP:journals/ijcia/HinzNMW18}.

\subsubsection{\underline{Tunability-based Measurements}}
While \textit{Variance-based} measurements are interested in the global effect of a knob, \textit{tunability-based} measurements focus on ``good'' regions of the space which are better than the default configuration.
It can be used directly to determine the necessity of tuning a knob from the given default value. 
We introduce two typical local algorithms that can measure the tunability -- ablation analysis~\cite{ DBLP:conf/aaai/BiedenkappLEHFH17} and SHAP~\cite{DBLP:conf/nips/LundbergL17}.

\noindent\textbf{Ablation Analysis~\cite{DBLP:journals/heuristics/FawcettH16}}
 selects the knob whose change contributes the most to improve the performance of configurations.
It identifies the difference between configurations by modifying each knob iteratively from its default value to the value of well-performing configurations and evaluating the performance change~\cite{DBLP:journals/heuristics/FawcettH16}. 
For speedup, the evaluations are replaced by cheap predictions obtained from surrogates, e.g., random forest.
Given a set of observations, we fit a surrogate and conduct ablation analysis between the default configuration and the better ones. 

\noindent\textbf{SHAP~\cite{DBLP:conf/nips/LundbergL17}} (i.e., SHapley Additive exPlanations)  is a unified framework to interpret the performance change derived from classic Shapley value estimation~\cite{DBLP:journals/kais/StrumbeljK14} in cooperative game theory.
The performance change is decomposed additively between knobs.
SHAP computes each knob's contribution (i.e., SHAP value) for pushing the default performance to the target one.
Given a set of observations, each knob's tunability is calculated by averaging its positive SHAP value (assuming a maximization problem).



\begin{table}[t]
\caption{Algorithms for Optimizers. \cmark means the optimizer has specific design for the case in the column and -- means the optimizer does not have such design.}\label{tab:algorithm}
\small
 \scalebox{0.95}[0.95]{

\begin{tabular}{|c|c|c|}
\hline
Algorithm  & High-dimensionality &  Heterogeneity \\ \hline
Vanilla BO & --                & --               \\ \hline
One-Hot BO & --                & \cmark \\ \hline
Mixed-Kernel BO~\cite{Klein2017RoBOA}       & --                   & \cmark                     \\ \hline
SMAC~\cite{DBLP:conf/lion/HutterHL11}       & \cmark                     & \cmark                     \\ \hline
TPE~\cite{DBLP:conf/nips/BergstraBBK11}        & --                  & \cmark                     \\ \hline
TurBO~\cite{DBLP:conf/nips/ErikssonPGTP19}      & \cmark                     & --                   \\ \hline
DDPG~\cite{DBLP:journals/corr/LillicrapHPHETS15}       & \cmark                     & --                   \\ \hline
GA~\cite{lessmann2005optimizing}     &  --                    &      \cmark             \\ \hline
\end{tabular}}
\end{table}

\subsection{Configuration Optimization}
In the \textit{configuration optimization} module, an optimizer suggests promising configurations and updates its model based on the evaluation results iteratively. 
We introduce six state-of-the-art optimizers used by the database tuning systems or from the HPO community. 
We also summarize their designs in Table \ref{tab:algorithm}. 


\noindent\textbf{Vanilla BO} denotes the BO-based optimizer that adopts vanilla GP as its surrogate model.
Vanilla GP is widely used for objective function modeling in database configuration systems \cite{DBLP:journals/pvldb/DuanTB09, 
DBLP:conf/kdd/FekryCPRH20}, due to its expressiveness, well-calibrated uncertainty estimates and closed-form computability of the predictive distribution~\cite{automl}. 
A GP $\mathcal{N}(\mu(\boldsymbol{\theta}), \sigma^2(\boldsymbol{\theta}))$ is fully specified by mean $\mu(\boldsymbol{\theta})$ and a variance function $\sigma^2(\boldsymbol{\theta})$, which is expressed as:
\begin{equation}
\small
\begin{aligned}
    \mu(\boldsymbol{\theta}) &= k^T \left(K +\sigma^{2}I\right)^{-1} y, \\
    \sigma^2(\boldsymbol{\theta}) &= k(\boldsymbol{\theta}, \boldsymbol{\theta}) - k^T {\left(K+\sigma^{2}I\right)}^{-1} k,
\end{aligned}\label{equ:gp}
\end{equation}
where $k=[k(\boldsymbol{\theta_1},\boldsymbol{\theta}),...,k(\boldsymbol{\theta_n},\boldsymbol{\theta})]^T$ and K is the covariance matrix whose
$(i, j)$-th entry is $K_{i,j}=k(\boldsymbol{\theta_i},\boldsymbol{\theta_j})$. 
The kernel $k(\boldsymbol{\theta}, \boldsymbol{\theta}^{'})$ models the overall smoothness of the target function.
For Vanilla BO, the kernel function (e.g., RBF kernel in OtterTune) is calculated based on the Euclidean distance between two configurations, assuming the natural ordering property and continuity of configuration space.

\noindent\textbf{One-hot BO} denotes the BO-based optimizer that adopts one hot-encoding  for categorical variables, as original GPs assume continuous input variables. 
Specifically, each categorical feature with $k$ possible values is converted into $k$ binary features.

\noindent\textbf{Mixed-kernel BO~\cite{Klein2017RoBOA}} denotes the BO-based optimizer that adopts a mixed-kernel GP as its surrogate model.
It uses Matérn kernel for continuous knobs, Hamming kernel for categorical knobs (also one-hit encoded), and calculates their product.
Mátern kernel is a continuous kernel that generalizes the RBF with a smoothing parameter.
Hamming kernel is based on hamming distance, which is suitable to measure the distance between categorical variables.
The mixed-kernel BO has shown promising performance over the heterogeneous configuration space empirically ~\cite{Klein2017RoBOA}.



\noindent\textbf{SMAC~\cite{DBLP:conf/lion/HutterHL11}} 
( i.e., Sequential Model-based Algorithm Configuration) adopts a random forest based surrogate, which is known to perform well for high-dimensional and categorical input~\cite{DBLP:journals/ml/Breiman01}.
SMAC assumes a Gaussian model $N(y|\hat{\mu}, \hat{\sigma}^2)$, where the $\hat{\mu}$ and $\hat{\sigma}^2$ are the mean and variance of the  random forest.
SMAC supports all types of variables, including continuous, discrete, and categorical features. It has been the best-performing optimizer in Auto-WEKA~\cite{DBLP:conf/kdd/ThorntonHHL13}.

\noindent\textbf{TPE~\cite{DBLP:conf/nips/BergstraBBK11}}
( i.e., Tree-structured Parzen estimator) is a non-standard Bayesian optimization algorithm.
While GP and SMAC modeling the probability $p(y|\boldsymbol{\theta}, H)$ directly, TPE models $p(\boldsymbol{\theta} |y, H)$ by tree-structured Parzen density estimators.
TPE describes the configuration space by a generative process and supports categorical features. 
TPE has been used successfully in several papers~\cite{DBLP:conf/icml/BergstraYC13, DBLP:conf/kdd/ThorntonHHL13, DBLP:journals/corr/BergstraC13}.



\noindent\textbf{TuRBO~\cite{DBLP:conf/nips/ErikssonPGTP19}} (i.e.,
Trust-Region BO) proposes a local strategy for global optimization using independent surrogate models. 
These surrogates allow for local  modeling of the objective function and do not suffer from over-exploration. 
To optimize globally, TuRBO leverages a multi-armed bandit strategy to select a promising suggestion from local models.
TuRBO exhibits promising performances  when solving high dimensional optimization problems~\cite{DBLP:conf/nips/ErikssonPGTP19}. 

\noindent\textbf{DDPG~\cite{DBLP:journals/corr/LillicrapHPHETS15}} denotes the Deep Deterministic Policy Gradient (DDPG) algorithm that is adopted to learn the configuration tuning policy for DBMS.
It has been successfully adopted by CDBTune and Qtune.
While other reinforcement learning algorithms, such as Deep-Q learning~\cite{DBLP:journals/access/MaglogiannisNSM18}, are limited to setting a knob from a finite set of predefined values, DDPG can work over a continuous action space, setting a knob to any value within a range.
DDPG consists of two neural networks: the actor that chooses an action (i.e., configuration) based on the input states, and the critic that evaluates the selected action based on the reward. 
In other words, the actor decides how to suggest a configuration, and the critic provides feedback on the suggestion to guide the actor.

\noindent\textbf{GA~\cite{lessmann2005optimizing}}
(i.e., Genetic Algorithm) is a meta-heuristic inspired by the process of natural selection. 
In a genetic algorithm, a population of candidate solutions to an optimization problem is evolved toward better solutions, iteratively.
In each iteration, the fitness of each solution, which is usually the value of the objective function, is evaluated.
The candidates with higher fitness 
will have more chance to be selected, and their features (i.e., encoded configuration) will be recombined and mutated to form new candidates for the next iteration.
Genetic algorithms are simple yet effective and naturally support categorical features. 
They have been applied to various problems, including hyper-parameter optimization~\cite{young2015optimizing,real2017large}.


\subsection{Knowledge Transfer}\label{sec:men-transfer}
The \textit{knowledge transfer} module is designed to accelerate the target tuning task by leveraging the experience from historical tuning tasks.
We introduce three knowledge transfer frameworks -- workload mapping, RGPE, and fine-tuning.

\noindent\textbf{Workload Mapping} is proposed by OtterTune.
It matches the target workload to the most similar historical one based on the absolute distances of database metrics and reuses the historical observations from the similar workload. 
This strategy can be adopted by any BO-based optimizers assuming a Gaussian model. 

\noindent\textbf{RGPE~\cite{feurer2018scalable}} (i.e., ranking-weighted Gaussian process ensemble ) is an ensemble model for BO-based optimizers.
 RGPE combines similar base GP models of historical tasks via distinguishable weights. 
 The weights are assigned using relative ranking loss to generalize across  different workloads and various hardware environments.
The ensemble manner avoids the poor scaling that comes with fitting a single Gaussian process model with all the  observations from similar tasks.
RGPE is  adopted in ResTune to accelerate the tuning process of the target tasks.

\noindent\textbf{Fine-tune} is used in the RL-based optimizers~\cite{DBLP:conf/sigmod/ZhangLZLXCXWCLR19,DBLP:journals/pvldb/LiZLG19}.
For example, CDBTune and Qtune could pre-train a basic DDPG model by replaying historical workloads.
During the later practical use of the tuner, the tuner continuously gains feedback information of its recommendation for each user tuning request. 
It uses the feedback data to update its model by gradient descent (fine-tune).
Such a process is claimed to help the optimizer adapt to different workloads with fewer observations instead of training  from scratch.

\section{General Setup of Evaluation}\label{sec4}


\begin{table}[t]
\caption{Profile information for workloads.}\label{wkl}
\small
\begin{tabular}{ccccc}
\hline
Workload & Class  & Size & Table & Read-Only Txns \\ \hline
JOB               & Analytical      & 9.3G          & 21             & 100.0\%        \\
SYSBENCH          & Transactional   & 24.8G         & 150            & 43.0\%         \\
TPC-C             & Transactional   & 17.8G         & 9              & 8.0\%          \\
SEATS             & Transactional   & 12.7G         & 10             & 45.0\%         \\
Smallbank         & Transactional   & 2.4G          & 3              & 15.0\%         \\
TATP              & Transactional   & 6.3G          & 4              & 40.0\%         \\
Voter             & Transactional   & 0.06M         & 3              & 0.0\%          \\
Twitter           & Web-Oriented    & 7.9G          & 5              & 0.9\%          \\
SIBench           & Feature Testing & 0.5M          & 1              & 50\%           \\ \hline
\end{tabular}
\end{table}



Our study conducts experimental evaluation for the three key modules of configuration tuning systems: \textit{knob selection, configuration optimization, knowledge transfer}. 
The evaluation is driven by the practical questions we encountered when tuning the databases:

\noindent\textbf{Q1:} How to determine the tuning knobs? (i.e., Which importance measurements to use and how many knobs to tune?)\\
\noindent\textbf{Q2:} Which optimizer is the winner given different scenarios?\\
\noindent\textbf{Q3:} Can we transfer knowledge to speed up the target tuning task?\\
We describe the general setup of the evaluation in this section and leave the specific setting (e.g., procedure, metrics) when answering the corresponding questions.  

\subsection{Hardware, Workloads and Tuning Setting}
\noindent\textbf{Hardware.}
We conduct our experiments on cloud servers.
Each experiment consists of two instances.
The first instance is used for the tuning server, deployed with 4 vCPUs and 8 GB RAM.
The second instance is used for the target DBMS deployment with 8 vCPUs and 16 GB RAM.
We adopt version 5.7 of RDS MySQL.
And we use the MySQL default configuration as the default in our experiments except that we set a larger and reasonable buffer pool size (60\% of the instance's memory~\cite{MySQLBuffer}), since the default size (128 MB) in MySQL document largely limits the database performance.

\noindent\textbf{Workloads.}
To answer \textbf{Q1} and \textbf{Q2}, we use an analytical workload JOB and a transactional workload SYSBENCH.
JOB contains 113 multi-joint queries with realistic and complex joins~\cite{DBLP:journals/pvldb/LeisGMBK015}.
SYSBENCH is a multi-threaded benchmark frequently used for database systems.
To answer \textbf{Q3}, we use three OLTP workloads (SYSBENCH, TPC-C and Twitter) as target workloads, which have been adopted in previous studies~\cite{DBLP:conf/sigmod/AkenPGZ17,DBLP:conf/sigmod/ZhangLZLXCXWCLR19, DBLP:conf/sigmod/ZhangWCJT0Z021}.
And we construct diverse source workloads, including  Smallbank, SIbench, Voter, Seats, and TATP.
They cover various size and read-write ratios, as shown in  Table \ref{wkl}.

\noindent\textbf{Tuning Setting.}\label{sec:exp-tuning}
There are 197 configuration knobs in MYSQL 5.7, except the knobs that do not make sense to tune (e.g., path names of where the DBMS stores files).
According to the proprieties of black-box optimization~\cite{Eggensperger2013TowardsAE}, we
choose three configuration spaces with different sizes: small,
medium, and large, where we tune the most important 5, 20, and 197 (all) knobs ranked by the importance measurement, respectively.
we run three tuning sessions per optimizer and report the median and quartiles of the best performances.
Each tuning session is comprised of 200 iterations without specification.
We restart the DBMS each iteration since the change of many knobs requires the restart.
And then, we conduct a three-minute stress testing, replaying the given workload.
For OLTP workload, we use throughput as maximization objective, and for OLAP workload, we use the 95\% quantile latency as minimization objective.
For a failed configuration (i.e., the one causing DBMS crash or unable to start), we set the result for that iteration to be the worst performance ever seen in order to avoid the scaling problem~\cite{DBLP:journals/pvldb/AkenYBFZBP21}. 
Following iTuned and OtterTune, we initialize each tuning session for BO-based optimizers  by executing 10 configurations generated by Latin Hypercube Sampling (LHS)~\cite{DBLP:conf/wsc/McKay92}. 
When reporting the results, we calculate the performance improvement against the default configuration.

\subsection{Implementation}\label{sec:imple}
\noindent\textbf{Importance measurements.}
We compare the five importance measurements listed in Table \ref{tab:knobs}. 
For Lasso~\cite{tibshirani1996regression}, we adopt the implementation in OtterTune, which includes second-degree polynomial features to enable Lasso to detect interactions between pairs of knobs. 
For ablation analysis, we use random forest implemented by scikit-learn~\cite{DBLP:journals/jmlr/PedregosaVGMTGBPWDVPCBPD11} as the surrogate.
For fANOVA~\cite{DBLP:conf/icml/HutterHL14}, we adopt its official library~\cite{fANOVA}.
As for SHAP~\cite{DBLP:conf/nips/LundbergL17}, we adopt the implementation released by the authors except that we use the given default as the base configuration when calculating performance gain.

\noindent\textbf{Optimizers.}
We compare the optimizers in Table \ref{tab:algorithm}.
For Vanilla BO and one-hot BO, we use a similar design with OtterTune~\cite{DBLP:conf/sigmod/AkenPGZ17}: Gaussian process as the surrogate and Expected Improvement  as the acquisition function. 
We implement they via Botorch library~\cite{DBLP:conf/nips/BalandatKJDLWB20}.
We adopt the OpenBox~\cite{DBLP:conf/kdd/LiSZCJLJG0Y0021}'s implementation for mixed-kernel BO and GA.
For SMAC and TurBO, we adopt the implementation released by the authors~\cite{DBLP:conf/lion/HutterHL11, DBLP:conf/nips/ErikssonPGTP19}.
We implement DDPG via PyTorch ~\cite{DBLP:conf/nips/PaszkeGMLBCKLGA19} with the same neural network architecture as CDBTune~\cite{DBLP:conf/sigmod/ZhangLZLXCXWCLR19}.

\noindent\textbf{Knowledge transfer frameworks.}
For workload mapping, we implement the methodology in OtterTune.
For RGPE~\cite{feurer2018scalable}, we adopt the implementation in OpenBox~\cite{DBLP:conf/kdd/LiSZCJLJG0Y0021}.
We fine-tune DDPG's pre-trained models' weight  when obtaining the target observations.


\section{How to determine tuning knobs?}\label{sec5}
In order to conduct  configuration tuning, we need to choose tuning knobs that decide the configuration space.
Specifically, it connotes the following two  questions: 

\noindent\textbf{Q5.2:} Which importance measurement to use to evaluate the importance of configuration knobs?

\noindent\textbf{Q5.3:} Given the importance measurement, how many knobs to choose for tuning?


\subsection{Setup}\label{sec:exp1-setup}

To conduct knob selection, we first collect 6250 samples (i.e., observations) respectively for the two workloads SYSBENCH and JOB via Latin Hypercube Sampling (LHS)~\cite{DBLP:conf/wsc/McKay92}.  
LHS attempts to distribute sample points evenly across all possible values over the 197-dimensional space.
Then, we apply different importance measurements to generate knob importance rankings and  compare the tuning performances on each importance ranking.
We perform configuration optimization on small spaces (top-5 knob sets) and medium spaces (top-20 knob sets), using the eight optimizers in Table \ref{tab:algorithm}.
We collect the performance rankings of the importance measurements in terms of the best configuration found by each optimizer.
To further understand the performances of different important measurements, we conduct sensitivity analysis varying the number of training samples to compare their stability and model accuracy.
Finally, we vary the number of tuning knobs to analyze the effect of configuration space with different sizes.


\subsection{Which importance measurement to use?}\label{sec:exp1-knob}

\begin{figure}[t]
    \centering
    \includegraphics[width=\columnwidth]{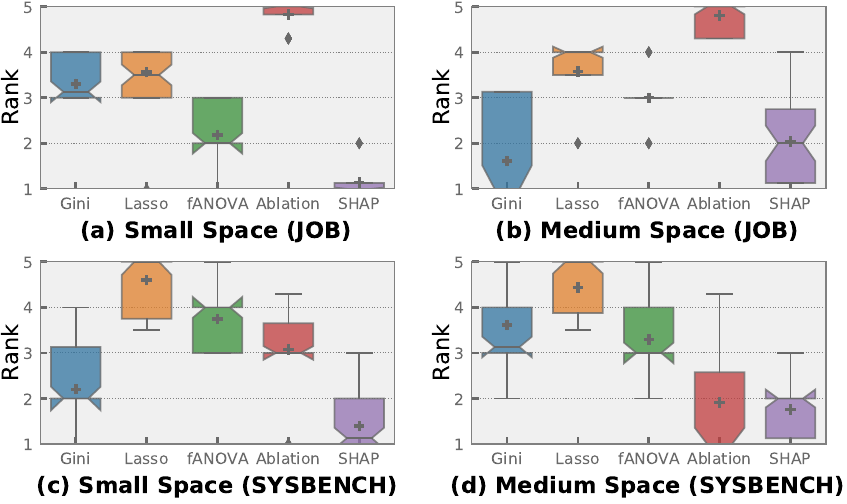}
    \caption{Performance ranking of importance measurements. (Notch denotes  medium and plus sign denote mean.)}
    \label{fig:exp1}
\end{figure}

\begin{table}[t]
\small
\setlength\tabcolsep{4pt}
\caption{Overall performance ranking (bold values are best).}\label{tab:imp-performance}
\begin{tabular}{cccccc}
\hline
Measurement  & Gini  & Lasso & fANOVA & Ablation Analysis & SHAP  \\ \hline
Average Ranking  &  2.62 & 4.18   & 3.06      & 3.45     & \textbf{1.67} \\ \hline
\end{tabular}
\end{table}

\begin{figure}[t]
    \centering
    \includegraphics[width=\columnwidth]{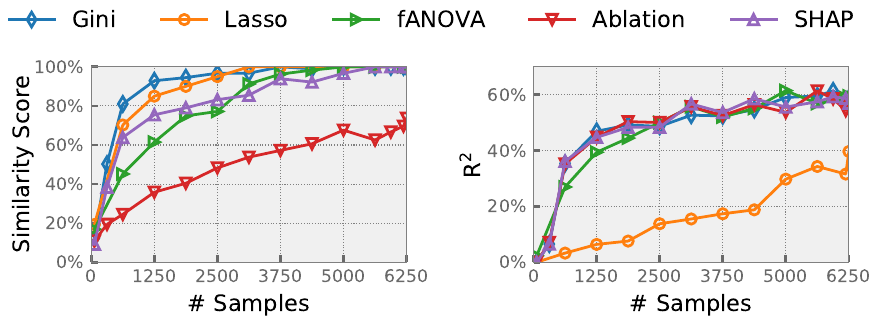}
    \caption{Sensitivity analysis for importance measurements.}\label{fig:exp1-sens}
\end{figure}


Figure \ref{fig:exp1} presents the performance ranking over the knob sets generated by different importance measurements, and Table \ref{tab:imp-performance} summarizes their overall performance ranking. 
We observe that the tunability-based method SHAP achieves the best performance.
This is due to that SHAP recommends the knobs worthy tuning.
When changing a knob's value from the default only leads to the downgrade of database performance, SHAP will recommend not to tune the knob, while variance-based measurements will consider this knob to have a large impact on the performance and need tuning.
The default values of database knobs are designed to be robust and can be a good start. 
Therefore, the variance-based measurements will be less effective. 
Ablation analysis yields the second last overall performance since it largely depends on the high-quality training samples better than the default and its performances are unstable as shown in Figure \ref{fig:exp1}.  
Among the variance-based measurements, Gini score performs the best overall, while Lasso tends to yield the worst improvement.
Lasso assumes a linear or quadratic configuration space, while in reality there exist complex dependencies from configurations to database performance. 


To further understand the different performance of importance measurements, we conduct sensitivity analysis on the number of training samples for SYSBENCH workload as shown in Figure~\ref{fig:exp1-sens}. 
The samples are randomly chosen from the 6250 samples, and the final result is the average of 10 executions for each importance measurement. 
The y-axis in the left figure is the similarity score (intersection-over-union index~\cite{DBLP:journals/ipm/Heine73}) of the top-5 important knobs ranked using the subset of training samples against that of the baseline (6250 samples). 
A larger similarity score indicates that the importance measurement is more stable since the measurement can find the final important knobs with fewer observations.
The right figure plots the Coefficient of Determination ~\cite{DBLP:journals/technometrics/Gunst99} (i.e., $R^2$) on the validation set.
A larger $R^2$ indicates that the surrogate  can better model the relationship between configurations and database performance.
We have that Lasso fails to model this relationship, while it is very stable.
Ablation analysis has the lowest stability, and its calculation highly depends on the high-quality samples.
Gini score has the highest similarity score, indicating its sample efficiency. 
In the meantime, SHAP has a similarity score comparable to the variance-based measurements. 
Considering that SHAP has the best overall performance and comparable stability, we use SHAP as the default measurement for the remainder of our study.


\begin{figure}[t]
    \centering
    \includegraphics[width=\columnwidth]{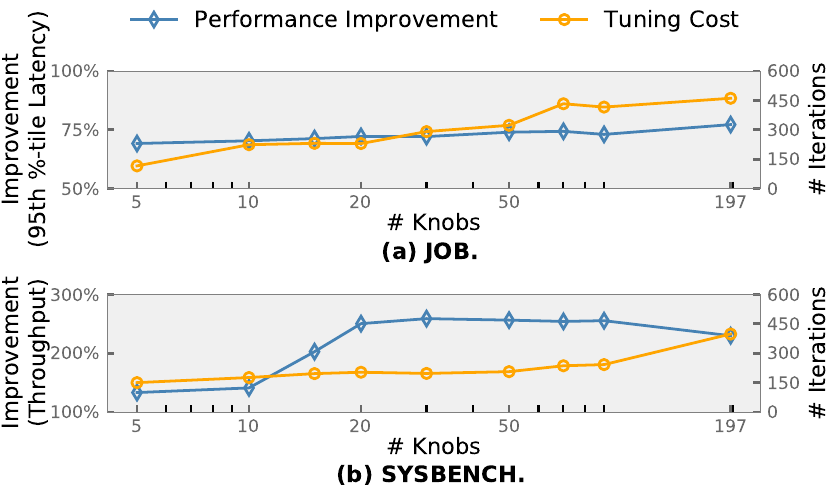}
    \caption{Performance improvement and tuning cost when increasing the number of tuning knobs.}\label{fig:knob-number}
    \label{fig:exp2}
     \vspace{-1em}
\end{figure}
\subsection{How many knobs to choose for tuning?}


We further conduct experiments to explore the effect of different numbers of tuning knobs. 
To measure the effect, we observe the tuning performance of vanilla BO over knob sets of different sizes ranked by SHAP.
For each set, we conduct tuning for 600 iterations on SYSBENCH and JOB.
We report the maximum performance improvement and the corresponding tuning costs (i.e., the iterations needed to find the configurations with the maximum improvement).

As shown in Figure \ref{fig:knob-number}, for JOB, the improvement is relatively stable, but the tuning cost increases when increasing the number of knobs. 
For SYSBENCH, the improvement is negligible at the beginning because a small number of tuning knobs have little impact on the performance. 
As the number of knobs grows from 5 to 20, the improvement increases from 133.10\% to 250.33\% since a larger configuration space leads to more tuning opportunities.
Afterward, the improvement decreases as the tuning complexity increases.
We conclude that it is better to tune the top-5 knobs for JOB and to tune the top-20 knobs for SYSBENCH.
There is a trade-off between performance improvement and tuning cost.
 Given more knobs to be tuned, the tuned performance is better, and more tuning iterations are required to achieve it.
With a limited tuning budget, it is vital to set the number of tuning knobs to an appropriate value since a small set of knobs only has a minor impact on the database performance, and a large set would introduce excessive tuning cost.

\begin{figure}[t]
    \centering
    \includegraphics{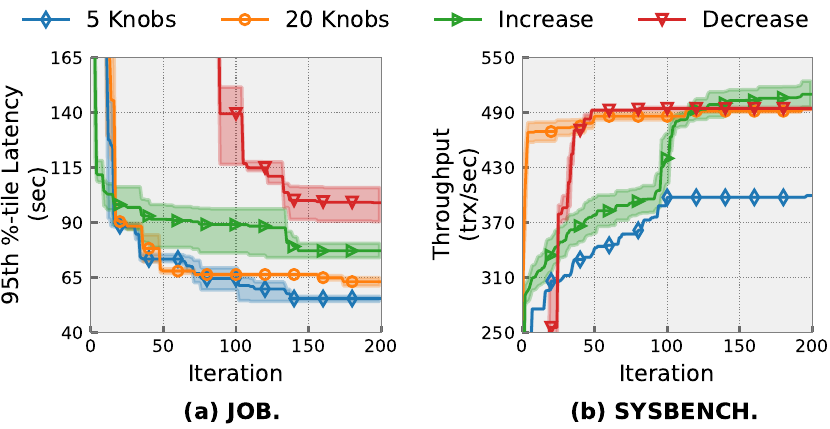}
    \caption{Best performance of incremental knob selection methods over iteration (for JOB, bottom left is better; for SYSBENCH, top left is better). We use top-5 knobs tuning and the top-20 knobs tuning as baselines. }
    \label{fig:exp_inc}
\end{figure}

\noindent\textbf{Incremental Knob Selection.}\label{sec:exp1-incremental}
Previously, we have enumerated the number of tuning knobs and conducted extensive tuning experiments. 
Using such a procedure to determine the configuration space is not practical in production due to the high costs. 
As discussed in Section \ref{sec2.3}, there are two incremental tuning heuristics to determine the number of knobs: (1) increasing the number of knobs, proposed in OtterTune~\cite{DBLP:conf/sigmod/AkenPGZ17}, and (2) decreasing the number of knobs, proposed in Tuneful~\cite{DBLP:conf/kdd/FekryCPRH20}. 
We implement the two methods.
For increasing the number of knobs, we begin  with tuning the top
four knobs and add two knobs  every four iterations.
For decreasing the number of knobs, we  begin with tuning all the knobs and remove 40\% knobs every 20 iterations.
Figure \ref{fig:exp_inc} presents the results.
For JOB, neither increasing nor decreasing the number of knobs surpass tuning the fixed 5 knobs.  In contrast, for SYSBENCH, increasing the number of knobs has better performance, but decreasing the number limits the potential performance gain.
As previously discussed, the number of important knobs for JOB is relatively small, thus the incremental knob selection on a larger space will not bring extra benefits. 
Instead, on SYSBENCH, the increasing method allows optimizers to explore a smaller configuration space of the most impacting knobs before expanding to a larger space. 


\subsection{Main Findings}

Our main findings of this section are summarized as follows:
\begin{itemize}[leftmargin=*]
    \item Given a limited tuning budget, tuning over the configuration space with all the knobs is inefficient.
    It is recommended to pre-select important knobs to prune the configuration space.
    
    \item Configuration spaces determined by different importance measurements will impact tuning performance significantly.
    \item SHAP is the best importance measurement based on our evaluation.
    Compared with traditional measurements (i.e., Lasso and Gini score), it achieves 38.02\% average performance improvement.
    When training samples are  limited, Gini score is also effective.
    
    \item When determining the number of tuning knobs, there is a trade-off between the performance improvement and tuning cost. Increasing/decreasing the number of tuning knobs has good performances in some cases. However, how to determine the number theoretically is still an open problem with research opportunities.

\end{itemize}

\section{Which optimizer is the winner?}\label{sec6}

In this section, we aim to find the best optimizer regarding the different sizes of configuration spaces. Furthermore, we construct two scenarios -  continuous configuration space and heterogeneous configuration space to validate the optimizers' support for heterogeneity. 
In addition, the algorithm overheads are also studied.

\subsection{Setup}\label{sec:exp2-setup}

We first evaluate the eight optimizers' performance over different spaces on workloads -- JOB and SYSBENCH.
To further validate the support for heterogeneity, we focus on the well-performing optimizers and construct a comparison experiment on JOB where we use the configuration space of top-20 integer knobs as a control group (continuous space) and the space of top-5 categorical knobs and top-15 integer knobs as a test group (heterogeneous space).
In addition, we measure the execution time of recommending a configuration (i.e., algorithm overhead) of different optimizers.


\begin{figure}[t]
    \centering
    \includegraphics[width=\columnwidth]{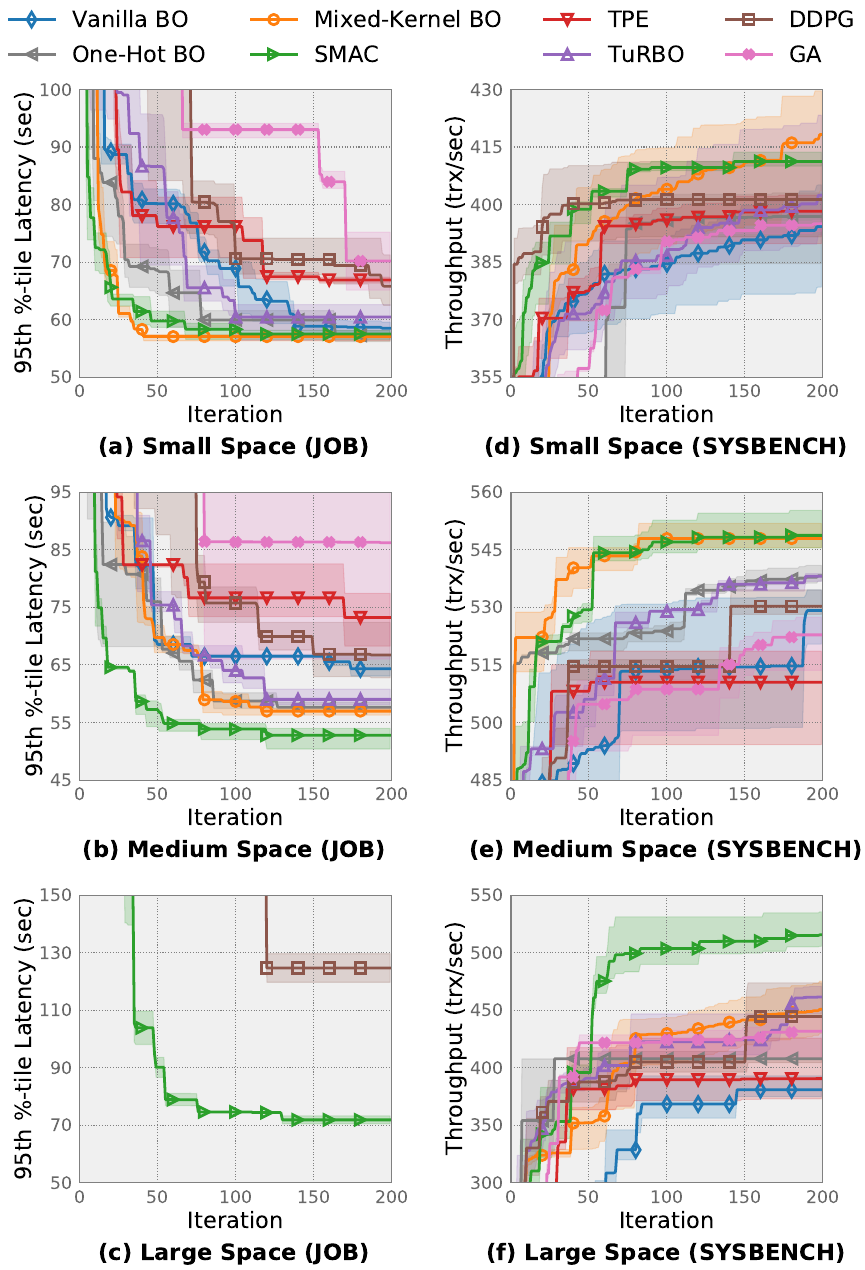}
    \caption{Best performance of optimizers over iteration (for JOB, bottom left is better; for SYSBENCH, top left is better).}
    \label{fig:exp3}
     \vspace{-1em}
\end{figure}

\subsection{Performance Comparison}\label{sec:exp2-performance}
\subsubsection{Performance over configuration spaces of different sizes}
Figure \ref{fig:exp3} presents the results, where each solid line denotes the mean of best performance across three runs, and shadows of the same color denote the quartile bar. We summarize the average ranking of optimizers in terms of the best performance they found in Table \ref{tab:optimizer-rank-1}.

We have that SMAC achieves the best overall performance, and TPE performs worst.
SMAC adopts random forest surrogate, which scales better to high dimensions and categorical input than other algorithms.
TPE fails to find the optimal configuration, and the main reason could be the lack of modeling the interactions between knobs~\cite{DBLP:conf/icml/BergstraYC13}.
It is almost certain that the optimal values of some knobs depend on the settings of others. 
For instance,  the tuning knobs \textit{tmp\_table\_size} and \textit{innodb\_thread\_concurreny} define
the maximum size of in-memory temporary tables and the maximum number of threads permitted.
Intuitively, the larger number of threads running, the more in-memory temporary tables created. 
The relationship can be modeled by the considered optimizers, while TPE does not.
In addition, the meta-heuristic method -- GA also performs poorly.


Over small and medium configuration spaces, SMAC and mixed-kernel BO exhibit outstanding performance. 
While both adopting the Gaussian Process, the BO-based optimizers  have distinct performance due to their different design with the categorical knobs (see the detailed analysis in \autoref{heter}).
One-hot BO performs better than vanilla BO, but is inferior to mixed-kernel BO.
In addition, DDPG has relatively bad performance on the small and medium spaces. 
It learns a mapping from internal metrics (state) to configuration (action). 
However, given a target workload, the optimal configuration is the same for any internal metrics, leading to the fact that action and state are not necessarily related~\cite{DBLP:conf/sigmod/ZhangWCJT0Z021}.
We also notice that the tuning cost of DDPG is constantly high due to the requirement of learning a great number of neural network parameters,  which is consistent with the existing evaluation~\cite{DBLP:journals/pvldb/AkenYBFZBP21}.



Over the large space of JOB, only SMAC and DDPG have found the configurations better than the default latency (about 200s) within 200 iterations.
The success of DDPG can be attributed to the good representation ability of the neural network to learn the high dimensional configuration space.
Over the large space of SYSBENCH, all methods have found improved configurations, among which SMAC still performs the best.
And TuRBO ranks the second-best since its local modeling strategy avoids the over-exploration in boundaries, especially in the high dimensional space.
The effectiveness of global GP methods (vanilla BO and mixed-kernel BO) further decreases when the number of tuning knobs increases.

\begin{table}[t]
\caption{Average ranking of optimizers in terms of the best performance. VBO, OHBO, MBO denote vanilla BO, one-hot BO, mixed-kernel BO respectively.  (bold values are the best.)}\label{tab:optimizer-rank-1}
\footnotesize

\setlength\tabcolsep{4.5pt}
\begin{tabular}{ccccccccc}

\hline
Optimizer     & VBO & OHBO & MBO & SMAC & TPE  & TuRBO & DDPG & GA\\ \hline
Small   & 5.33 & 4.00 & \textbf{2.17} & 3.33 & 5.83 & 3.83 & 5.00 & 6.50 \\
Medium  & 5.17 & 3.83 & 2.33 & \textbf{1.33} & 7.17 & 4.00 & 5.50 & 6.67 \\
Large   & 7.33 & 6.50 & 5.17 & \textbf{1.00} & 6.50 & 5.00 & 3.17 & 5.83 \\
\hline
\textbf{Overall} & 5.94 & 4.78 & 3.22 & \textbf{1.89} & 6.50 & 4.28 & 4.56 & 6.33 \\ \hline
\end{tabular}
\end{table}

\vspace{-1em}

\subsubsection{Comparison Experiment for Knobs Heterogeneity.}\label{heter}

\begin{figure}[t]
    \centering
    \includegraphics[width=\columnwidth]{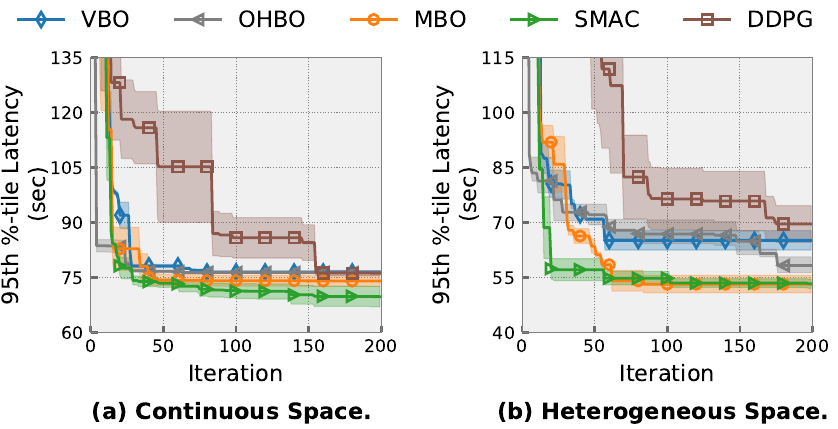}
    \caption{Comparison experiment for knobs heterogeneity (bottom left is better).}
    \label{fig:exp_space} 
\end{figure}

Figure \ref{fig:exp_space} demonstrates the performance of vanilla BO, one-hot BO, mixed-kernel BO, SMAC, and DDPG on continuous and heterogeneous spaces, respectively.
While the BO-based optimizers perform similarly on continuous space, they reach quite different performances on heterogeneous space. 
Mixed-kernel BO could find better configurations and with a faster convergence speed than the others. 
The reasons are as follows: 
vanilla BO  cannot handle categorical knobs since it assumes a partial order between the different values of a categorical knob~\cite{DBLP:conf/icml/WanNHRLO21}.
Although one-hot encoding converts categorical knobs into  binary ones, its RBF kernel struggles to capture the distance between categorical knobs.
Mixed-kernel BO adopts different kernels for the integer and categorical knobs, thus better measuring the distance in heterogeneous spaces.
In addition, SMAC performs well over the two spaces due to its random forest modeling, and DDPG has high tuning costs for finding good configurations.

\subsection{Algorithm Overhead Comparison}
\begin{figure}[t]
    \centering
    \includegraphics[width=\columnwidth]{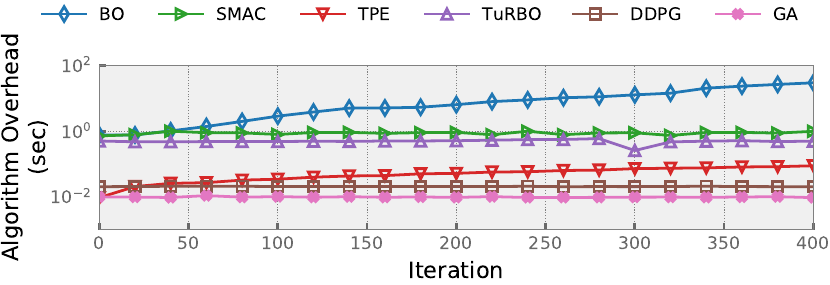}
    \caption{Algorithm overhead for different optimizers.}
    \label{fig:overhead}
     \vspace{-1em}
\end{figure}

Algorithm overhead is the execution time taken by an optimizer to generate the next configuration to evaluate per iteration, 
and does not include the evaluation time.
Precisely, it consists of the time for statistics collection, model fitting, and model probe~\cite{DBLP:conf/sigmod/KunjirB20}.
Figure \ref{fig:overhead} shows the statistics when tuning workload - JOB over medium configuration space.
GA has the lowest algorithm overhead. 
DDPG, SMAC, and TPE also have negligible overhead (< 1 second).
However, due to the cubic scaling behavior of GP, the overhead of BO-based methods become extraordinarily expensive as the number of iterations increases.
In particular, it takes > 10 seconds to select the next configuration after 200 iterations and > 1 minute after 400 iterations.
TuRBO's overhead is comparable to SMAC.
TuRBO uses a collection of simultaneous local GPs instead of a global GP and terminates unpromising GPs, which mitigates the scaling problem.


\subsection{Main findings}
Our main findings of this section are summarized as follows:
\vspace{-0.3em}
\begin{itemize}[leftmargin=*]
    \item SMAC has the best overall performance and could simultaneously handle the high-dimensionality and heterogeneity of configuration space.
    Compared with traditional optimizer (i.e., vanilla BO, DDPG), it achieves 21.17\% average performance improvement.
    \item TPE is worse than other optimizers in most cases since it struggles to model the interaction between knobs.
    
    \item DDPG has considerable tuning costs (i.e., more iterations) in small and medium configuration spaces due to its redundant MDP modeling and complexity of the neural network. Meanwhile, it has a relatively good performance in a large configuration space.
    
    \item On small and medium configuration spaces, SMAC and mixed-kernel  BO rank the top two, while on the large configuration space, SMAC, DDPG, and TuRBO all have good performance rankings.
    The effectiveness of global GP methods decreases as the number of tuning knobs increases.
    
    \item Mixed-kernel  BO outperforms other BO-based optimizers in heterogeneous  space due to its Hamming kernel measurement. 
    
    \item 
    Global GP-based optimizers (i.e., vanilla BO, one-hot BO and mixed-kernel BO) have cubic algorithm overhead.

\end{itemize}

\begin{table*}[t]
\footnotesize
\caption{Evaluation results for different transfer frameworks (the bold values are the best). We report speedup, performance enhancement (i.e., PE) against the base optimizers and the absolute performance ranking (i.e., APR).}\label{tab:transfer}
 \scalebox{0.95}[0.95]{

\begin{tabular}{c|ccc|ccc|ccc|ccc|ccc}
\hline
Transfer Framework & \multicolumn{6}{c|}{RGPE}                                        & \multicolumn{6}{c|}{Workload Mapping}                                                                             & \multicolumn{3}{c}{Fine-Tune} \\ \hline
Base Optimizer     & \multicolumn{3}{c|}{Mixed-Kernel BO} & \multicolumn{3}{c|}{SMAC} & \multicolumn{3}{c|}{Mixed-Kernel BO}                    & \multicolumn{3}{c|}{SMAC}                               & \multicolumn{3}{c}{DDPG}      \\
Metric             & Speedup      & PE    & APR     & Speedup  & PE & APR  & Speedup                                & PE & APR  & Speedup                                & PE & APR  & Speedup   & PE   & APR    \\ \hline
TPCC               & \textbf{98.28}        & \textbf{10.44\%}     & \textbf{1}       & 8.03     & 2.18\%  & 2    & $\times$ & -2.48\% & 4    & 0.35                                   & 2.43\%  & 3    & 1.71      & 3.75\%    & 4      \\
SYSBENCH           & \textbf{4.76}         & 0.53\%      & 4       & 0.78     & \textbf{13.32\%} & \textbf{1}    & $\times$ & -0.51\% & 5    & 3.08                                   & 2.23\%  & 2    & 0.93      & 4.59\%    & 3      \\
Twitter            & \textbf{28.42}        & 1.56\%      & \textbf{1}       & \textbf{28.42}    & 0.02\%  & 2    & $\times$ & -1.70\% & 5    & $\times$ & -0.12\% & 4    & 0.83      & \textbf{3.12\%}    & 3      \\
\hline
\textbf{Avg.}               & \textbf{51.52}        & 4.18\%      & 2       & 12.41    & \textbf{5.17\%}  & \textbf{1.67} & $\times$ & -1.56\% & 4.67 & 1.14                                   & 1.51\%  & 3.00 & 1.32      & 3.82\%    & 3.33   \\ \hline
\end{tabular}}
\vspace{-1em}
\end{table*}

\begin{table}[t]
\caption{Regression performance (bold values are the best).}\label{tbl:benchmark}
\setlength\tabcolsep{0.5pt}
\footnotesize
 \scalebox{0.95}[0.95]{

\begin{tabular}{c|cc|cc|cc|cc|cc|cc}
\hline
Model    & \multicolumn{2}{c|}{RF} & \multicolumn{2}{c|}{GB} & \multicolumn{2}{c|}{SVR} & \multicolumn{2}{c|}{NuSVR} & \multicolumn{2}{c|}{KNN} & \multicolumn{2}{c}{RR} \\
Metric   & RMSE      & $R^2$       & RMSE      & $R^2$       & RMSE       & $R^2$       & RMSE        & $R^2$        & RMSE       & $R^2$       & RMSE      & $R^2$       \\ \hline
SYSBENCH & \textbf{26.5}      & \textbf{93.0\%}      & 27.2      & 92.6\%      & 97.4       & 5.6\%       & 97.4        & 5.6\%        & 54.6       & 70.2\%      & 64.1      & 59.1\%      \\
JOB      & 11.8      & 97.4\%      & \textbf{11.1}      & \textbf{97.7\%}      & 41.7       & 67.9\%      & 41.7        & 67.9\%       & 27.5       & 86.0\%      & 52.3      & 49.5\%      \\ \hline
\end{tabular}}
\end{table}

\section{Can we transfer knowledge to speed up the target tuning task?}\label{sec:exp-transfer}

In the previous sections, different optimizers are compared from scratch without knowledge transfer.
In this section, we test whether we can utilize the historical data to speed up target tuning tasks and compare the applicability of different transfer frameworks.

\noindent\textbf{Baselines.}
The transfer learning framework is used to speed up  base optimizers. 
To narrow down the candidate baselines, we evaluate workload mapping and RGPE accelerating the best-performing BO-based optimizers -- SMAC and mixed-kernel BO.
Then we have five baselines:  Mapping (SMAC), Mapping (Mixed-Kernel BO), RGPE (SMAC), RGPE (Mixed-Kernel BO), and Fine-tune (DDPG).

\noindent\textbf{Metrics.}
We use three metrics to evaluate the performance of transfer frameworks: performance enhancement, speedup, and absolute performance.
The performance enhancement and speedup focus on whether the transfer framework can speed up the tuning process compared with non-transfer.
We denote the best performance within 200 iterations for the base optimizer without transfer as $f(x_{base}^{*})$ and the best performance with transfer as $f(x_{tran}^{*})$.
The performance enhancement (PE) is defined as 
\begin{equation}
\small
PE=\frac{f(x_{tran}^{*})-f(x_{base}^{*})}{f(x_{base}^{*})};
\end{equation}
and the speedup ($\eta$) is calculated as:
\begin{equation}
\small
\eta = \frac{steps \  to\ find \ x_{base}^{*} without \ transfer}{steps  \ to \ find \ config. better \ than \ x_{base}^{*} with \ transfer }. \end{equation}
The absolute performance focuses on the performance of the combination of base learner and transfer framework (e.g., the best performance Mapping (SMAC) achieved within 200 iterations).

\noindent\textbf{Main Procedure.}
We conduct experiments on three target workloads -- SYSBENCH, TPC-C, and Twitter.
As for the knobs we tune, we use SHAP to select top-20 impacting knobs across OLTP workloads  and hardware instances and more details can be found in our Appendix.
To gather historical tuning data, we  collect observations from five source workloads -- SEATS, Voter, TATP, Smallbank, and SIBench.
Since fine-tune (DDPG) relays on a pre-trained model, we pre-train DDPG's network 300 iterations on the five source workloads in turn.
We use DDPG's training observations as the historical data for workload mapping and RGPE frameworks.
Such a setting follows the evaluation of CDBTune for data fairness.
With the pre-trained models and source observations, we compare the five baselines and obtain absolute performances.
To calculate performance enhancement and speedup, we also run base optimizers on the target workloads without knowledge transfer. 

\subsection{Performance Comparison}

Table \ref{tab:transfer} shows the results.
If a baseline fails to find configurations better than $x_{base}^{*}$, we put ``$\times$'' in the speedup column.
We observe that fine-tune and workload mapping hinder the optimization (i.e., ``negative transfer'') in some cases where the speedup is smaller than one or the improvement is negative.
Workload mapping always maps a similar workload and combines its observations in the surrogate model together with the target observations, which may be problematic since the source workloads may not be entirely identical to the target one.  
RGPE solves this problem by assigning adaptive weights to source surrogates and utilizing them discriminately.
As for  fine-tune, the performance is not stable since the neural network may over-fit the source workloads, and fine-tuning the over-fitted network may be less efficient than training from scratch. 
Overall, the RGPE framework has the best performance. 
 RGPE (mixed-kernel BO) has shown impressive speedup  accelerating, which may come from the inferior performance of mixed-kernel BO compared with SMAC.
In terms of absolute performance, RGPE (SMAC) achieves the best overall performance.

\begin{figure}[t]
    \centering
    \includegraphics{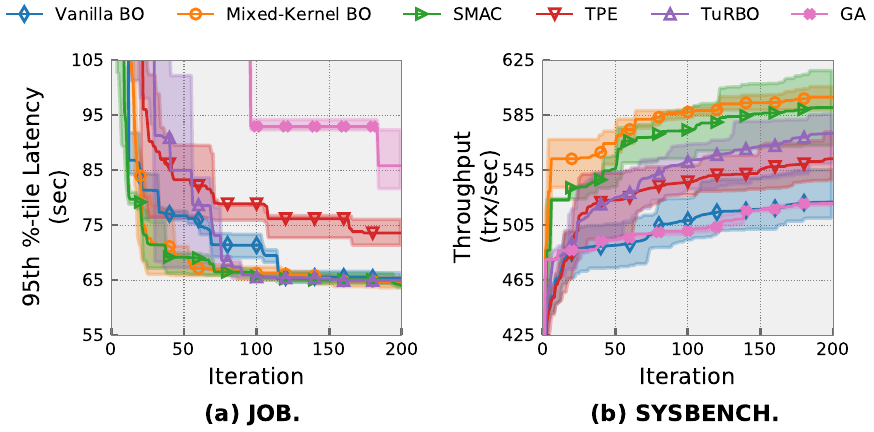}
    \caption{Tuning Performance over surrogate benchmark.}\label{fig:benchmark}
    \label{fig:exp-bench}
\end{figure}

\begin{figure}[t]
    \centering
    \includegraphics{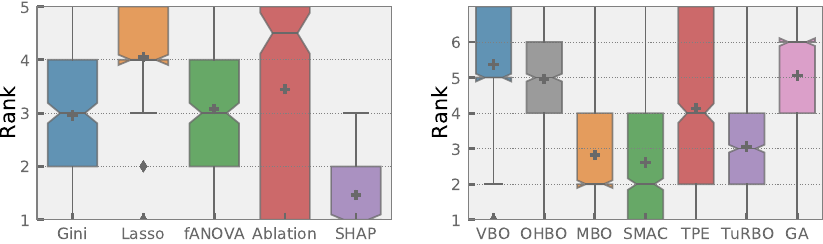}
    \caption{Performance ranking of importance measurements and optimizers evaluated by the tuning  benchmark. (Notch denotes  medium and plus sign denote mean.)}
    \label{fig:exp-bench2}
     \vspace{-1em}
\end{figure}

\section{Efficient database tuning benchmark via surrogates}\label{sec8}
As discussed previously, to ease the burden of evaluating tuning optimizers, we propose to benchmark the database tuning via surrogate models that approximate expensive evaluation through cheap and stable model predictions.  
A user can easily test optimizers by interacting with the surrogate models (i.e., input the configuration suggested by the optimizer and output the corresponding database performance).
We present the construction of the tuning benchmarks and the evaluation results based on the benchmarks.  

To construct the tuning benchmarks, we first collect extensive training samples and then select a regression model with high accuracy as the surrogate.
To collect the training data, we run existing database optimizers to densely sample high-performance regions of the configuration space and sample poorly-performing regions ~\cite{DBLP:conf/aaai/EggenspergerHHL15} via LHS. 
As for the regression model, we consider a broad range of commonly used models as candidates, including Random Forest (RF), Gradient Boosting (GB), Support Vector Regression (SVR), NuSVR, k-nearest-neighbours (KNN), Ridge Regression (RR).
We evaluate their performance via 10-fold cross-validation, and Table \ref{tbl:benchmark} presents the resulted mean squared error (RMSE) and  $R^2$.
We have that the two tree-based models, RF and GB, perform the best.
Since RFs are widely
used with simplicity, we adopt RF as the surrogate for the tuning benchmark.

We first focus on the small configuration space of JOB and medium space of SYSBENCH.
Figure \ref{fig:benchmark} depicts the best performance found by different optimizers using the tuning benchmark based on RFs.
We report means and quartiles across ten runs of each optimizer. 
We observe that our tuning benchmark yields evaluation results closer to the result in Figure \ref{fig:exp3} -- SMAC and mixed-kernel BO have the best overall performance.
In addition, the experiments on our tuning benchmarks are much faster.
For example, as previously mentioned, a single function evaluation on SYSBENCH workload requires at least 3 minutes, while a surrogate evaluation needs 0.08 seconds on average. 
When considering the algorithm overhead of optimizers, the previous  200-iteration experiment takes at least 10 hours, while the same experiment on the tuning benchmark takes about 2\textasciitilde4 minutes.
The tuning benchmark brings 150\textasciitilde311 $\times$ speedups.
We leave the benchmarking RL-based optimizers as future work, since it requires constructing a surrogate to learn the state transaction (i.e., internal metrics of DBMS).


In addition, using surrogates could enlarge the number of evaluation cases of tuning algorithms
To achieve this goal, we collect samples over the large (197-dimensional) configuration space and fit a surrogate.
In the previous evaluation, we conduct experiments over three configuration spaces since evaluating the enumeration of all  dimensions is expensive.
And when evaluating the optimizers, we fix the importance measurement to be SHAP.
With the surrogate benchmark, we could conduct extensive experiments over all the dimensions of configuration spaces (from top-1 to top-197) and all the importance measurements.
Then, we conduct 11820 experiments (197 knob sets $\times$  5 importance measurements $\times$ 6 optimizers $\times$  2 workloads) to benchmark the \textit{knob selection} and \textit{configuration optimization}  modules.
Since the \textit{knob selection}  aims to prune the configuration space, we report its performance on 20 knob sets (top-1 to top-20). 
And we report the performance of optimizers on the 197 knob sets.
As shown in Figure \ref{fig:exp-bench2}, SHAP remains the best importance measurement, and ablation study has significant performance variance.
SMAC performs the best, followed by mixed-kernel BO.
The evaluation results further validate our conclusions.

\section{Discussion}\label{sec9}
We summarize our answers to the motivating questions and discuss the research opportunities we draw from the evaluations.

\subsection{Answers to The Motivating Questions}

\vspace{0.3em}
\noindent\textbf{A1:} 
We recommend the tunability-based method -- SHAP as an importance measurement since it indicates the necessity of tuning a knob when the DBMS has relatively reasonable default knob values.
We can determine the number of knobs to tune by exhaustive and expensive enumeration.
How to determine the number theoretically with fewer evluations is still an open problem.
In practice, with a limited tuning budget, we could resort to incremental tuning. 

\vspace{0.3em}
\noindent\textbf{A2:} When comparing different optimizers, we need to consider the size and composition of configuration space.
SMAC and DDPG are recommended for high-dimensional configuration spaces, although it is better to first conduct knob selection to prune the configuration space.
SMAC is the winner that can handle the high high-dimensionality and heterogeneity of configuration space.

\vspace{0.3em}
\noindent\textbf{A3:} 
For RL-based optimizers, we can fine-tune its pre-trained model to adapt to the target workload.
However, we find that it might suffer from the negative transfer issue empirically.
For BO-based optimizers, RGPE exhibits excellent speedup and improvement since it avoids negative transfer via adaptive weight assignment.

In summary,  we have that using SHAP measurement to prune the unimportant knobs and adopting SMAC optimizer in the RGPE transfer framework could reach the best end-to-end performance. 

\subsection{Research Opportunities}


\noindent\textbf{An End-to-End Optimization for Designing Database Tuning Systems.}
The \textit{end-to-end optimization} can be viewed as optimizing over a joint search space, including the selection of importance measurements, knobs, optimizers, and transfer frameworks.
Because of the joint nature, the search space of the \textit{end-to-end optimization} is complex and huge.
In our evaluation, we decompose the joint space reasonably to narrow the search space.
Meanwhile, a class of methods in the HPO field treats the selection of algorithms as a new hyper-parameter to optimize. 
They optimize over the joint search space with probabilistic models~\cite{DBLP:conf/nips/FeurerKESBH15,DBLP:journals/jmlr/KotthoffTHHL17, DBLP:conf/aaai/LiJGSZ020, DBLP:conf/kdd/GaoYJL21,Li2021VolcanoMLSU}, which could be another research direction.

\noindent\textbf{Tuning Budget Allocation.}
How to allocate a limited tuning budget between different modules (e.g., knob selection and knobs optimization) is a problem that needs exploring~\cite{DBLP:journals/apjor/WangXH21,DBLP:journals/asc/FuXLH21}. 
For example, an accurate ranking of knobs can facilitate later  optimization but comes with the cost of collecting extensive training samples.
There is a trade-off between the budgets for sample collection and the latter optimization.
In addition, as discussed, a larger configuration space gives us more tuning opportunities but with higher tuning costs.
There still remains space when determining the number of tuning knobs wisely, given a limited tuning budget.

\section{Conclusion}\label{sec10}
Given emerging new designs and algorithms for configuration tuning systems of DBMS, we are curious about the best solution in different scenarios.   
In this paper, we decompose existing systems into three modules and comprehensively analyze and evaluate the corresponding intra-algorithms to construct optimal design ``paths'' in various scenarios.
Meanwhile, we identify the design trade-offs to suggest insightful principles and promising research directions to optimize the tuning systems.
In addition, we propose an efficient database tuning benchmark that reduces the evaluation overhead to a minimum, facilitating the evaluation and analysis for new algorithms with fewer costs. 
It is noted that we do not restrict our evaluation within the database community and extensively evaluate promising approaches from the HPO field.
Our evaluation demonstrates that such an out-of-the-box manner can further enhance the performance of database configuration tuning systems.

\bibliographystyle{ACM-Reference-Format}
\balance
\bibliography{main}


\newpage

\subtitle{[Supplemental Material]}
\setcounter{table}{0}
\setcounter{figure}{0}
\setcounter{section}{0}
\renewcommand{\thetable}{S\arabic{table}}
\renewcommand\thefigure{S\arabic{figure}}
\renewcommand\thesection{S\arabic{section}}
\renewcommand{\theHtable}{Supplement.\thetable}
\renewcommand{\theHfigure}{Supplement.\thefigure}
\renewcommand{\theHsection}{Supplement.\thesection}

\noindent\textbf{\Huge{Appendix}}
\vspace{1em}

\noindent\textbf{{\LARGE OUTLINE}}

\noindent This supplemental material is organized as follows.

\noindent\textbf{S1.} More background for evaluating configuration tuning systems.

\noindent\textbf{S2.}  Details about  intra-algorithms.

\noindent\textbf{S3.} Construction for database configuration tuning benchmark.

\noindent\textbf{S4.}  More details and results about the experiment.

\noindent\textbf{S5.}  Evaluations on PostgreSQL.

\noindent\textbf{S6.} Experimental environment and reproduction instructions.

\begin{figure*}[hb]
    \centering
    \scalebox{0.98}[0.98]{
    \includegraphics[width=1\linewidth]{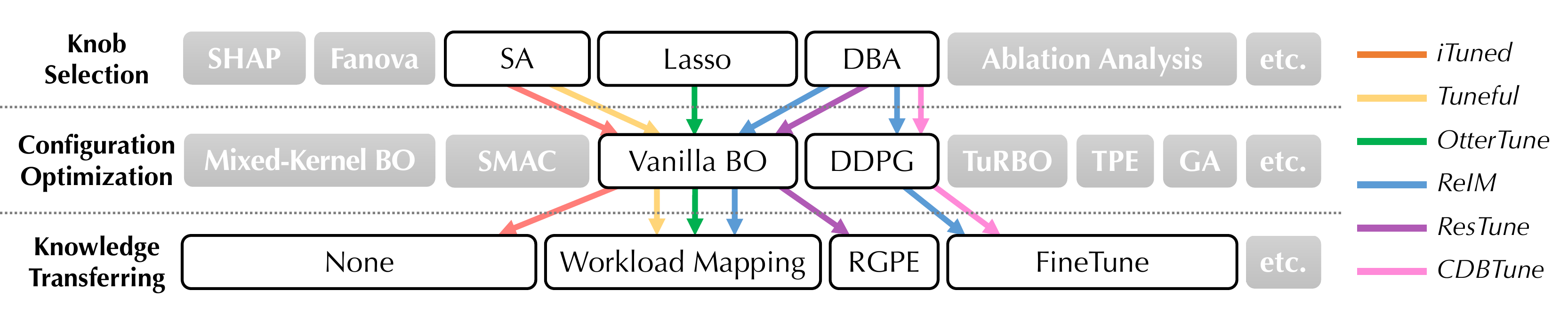}
    }
    \caption{Detailed Intra-algorithm Designs of Configuration Tuning Systems (The Full Picture): 
    Black boxes denote the algorithms adopted by existing database tuning systems (indicated by colored paths), and grey boxes denote the algorithms in the HPO field. SA denotes sensitivity analysis based on Gini score and GA denotes Genetic algorithm. 
    }
    \label{fig:ap-intro}
\end{figure*}

\section{More background for evaluating configuration tuning systems}\label{appendix}

The previous evaluation for database configuration tuning is limited to a subset of existing systems where the analysis and evaluation of intra-algorithm components are ignored. 
Instead, we identify three key modules of configuration tuning systems and conduct a thorough analysis and experimental evaluation from a micro perspective (i.e., evaluating every fine-grained algorithm).
Figure \ref{fig:ap-intro} presents the fine-grained algorithms adopted by existing database tuning systems or from the HPO field. 
When conducting database configuration tuning in practice, we have to chose a solution ``path'' across the three modules:  (1) \textit{knob selection}, (2) \textit{configuration optimization}, and (3) \textit{knowledge transfer}, as shown in the figure. 
Each \textit{knob selection} algorithm could determine an unique configuration space and can be ``linked'' to any of the \textit{configuration optimization} algorithms (i.e., optimizers).
And among the optimizers, all the BO-based optimizers assuming a Gaussian model (SMAC, vanilla BO, mixed-kernel BO, TurBO) can be ``linked'' to workload mapping or RGPE transfer frameworks.
And the DDPG algorithm is ``linked'' to fine-tune framework.
We have noted that existing systems only cover a part of the possible solutions and it remains unclear to identify the best “path” for database configuration tuning.
We evaluate all the fine-grained algorithms listed in Figure \ref{fig:ap-intro} and carefully decompose the search (evaluation) space to identify the best ``path'' in various scenarios.

\section{Details about intra-algorithms}
In this section, we present details about the intra-algorithms which we describe on a high level in the paper due to space constraints.

\subsection{Ablation Analysis}
Ablation analysis \cite{DBLP:conf/aaai/BiedenkappLEHFH17} selects the features whose changes contribute
the most to improve the performance of configurations.
We now describe how ablation analysis quantifies the performance change due to a certain feature's change.
Given a default configuration $\boldsymbol{\theta}_{default}$ and a target configuration $\boldsymbol{\theta}_{target}$ (usually a better one), ablation analysis first computes the feature differences  $\Delta(\boldsymbol{\theta}_{default},\boldsymbol{\theta}_{target} )$ between the default and target configurations.
Next, an ablation path $\boldsymbol{\theta}_{default}, \boldsymbol{\theta}_1, \boldsymbol{\theta}_2, \dots , \boldsymbol{\theta}_{target}$ is iteratively constructed.
In each iteration $i$ with previous ablation path configuration $\boldsymbol{\theta}_{i-1}$, we consider all remaining feature changes $\delta \in \Delta(\boldsymbol{\theta}_{default},\boldsymbol{\theta}_{target} )$ and apply the change to the previous ablation path configuration and obtain the candidate $\boldsymbol{\theta}_{i-1}[\delta]$.
Each parameter change $\delta$ is a modification of one feature from its value in $\boldsymbol{\theta}_{i-1}$ to its value in $\boldsymbol{\theta}_{target}$, along with any other feature modifications that may be necessary due to conditionally constraints in $\boldsymbol{\Theta}$. 
The next configuration on the ablation path $\boldsymbol{\theta}_{i}$ is the candidate $\boldsymbol{\theta}_{i-1}[\delta]$ with the best objective performance $f$. 
The performance evaluation is approximated via surrogate for efficiency reasons.
The order that the feature is changed is the importance rank from the ablation analysis.
Given a set of observations, we conduct the ablation analysis between the default configuration and the configurations with better performance than the default in the observation set (i.e., target configurations). 
For each feature, we use the average rank from each ablation path as the final ranking of importance.

\subsection{SHAP}
 SHAP~\cite{DBLP:conf/nips/LundbergL17} (SHapley Additive exPlanation) uses Shapley values of a conditional expectation function of the original model.
 SHAP values attribute to each feature the change in the expected model prediction when conditioning on that feature. 
 They explain how to get from the base value that would be predicted if we did not know any features to the current output.  
 When the model is non-linear or the input features are not independent, the order in which features are added to the expectation matters, and the  SHAP values arise from averaging the contributing values across all possible orderings~\cite{DBLP:conf/nips/LundbergL17}.
The exact computation of SHAP values is challenging, which can be estimated by Shapley sampling values method~\cite{DBLP:journals/kais/StrumbeljK14} or Kernel SHAP method~\cite{DBLP:conf/nips/LundbergL17}.

\subsection{SMAC}
SMAC~\cite{DBLP:conf/lion/HutterHL11} constructs a random forest as a set of regression trees, each of which is built on n data points randomly sampled with repetitions from the entire training data set.
It computes the random forest’s predictive mean  $\hat{\mu}(\boldsymbol{\theta})$ and variance $\hat{\sigma}^2(\boldsymbol{\theta})$ for a new configuration $\boldsymbol{\theta}$ as the empirical
mean and variance of the Gaussian  distribution .
SMAC uses the random forest model to select a list of promising parameter configurations.
To quantify how promising a configuration $\boldsymbol{\theta}$ is, it uses the model’s predictive distribution for $\boldsymbol{\theta}$  to compute its expected positive improvement $EI(\boldsymbol{\theta} )$ ~\cite{DBLP:journals/jgo/JonesSW98} over the best configuration seen so far.
$EI(\boldsymbol{\theta} )$ is large for configurations $\boldsymbol{\theta}$ with high predicted performance and for those with high predicted uncertainty; thereby, it offers an automatic trade-off between exploitation (focusing on known good parts of the space) and exploration (gathering more information in unknown parts of the space). 
To gather a set of promising configurations with low computational overhead, SMAC performs a simple multi-start local search and considers all resulting configurations with locally maximal $EI$.

\subsection{RGPE}
RGPE~\cite{feurer2018scalable} is a scalable meta-learning framework to accelerate BO-based optimizer.
First, for each previous tuning task $T_i$, it trains a base Gaussian process (GP) model $M_i$ on the corresponding observations from $H_i$. 
Then it builds a surrogate model $M_{meta}$ combine the base GP models, instead of the original surrogate $M_T$ fitted on the observations $H_T$ of the target task only.
The prediction of $M_{meta}$ at point $\boldsymbol{\theta}$ is given by:
\begin{equation}
    y \sim N(\sum_{i}{w_i\mu_i(\boldsymbol{\theta})}, \sum_{i}{w_i\sigma^2_i(\boldsymbol{\theta})}),
\end{equation}
where $w_i$ is the weight of base surrogate $M_i$, and $\mu_i$ and $\sigma^2_i$ are the
predictive mean and variance of the base surrogate $M_i$.
The weight $w_i$ reflects the similarity between the previous task and the current task. Therefore, $M_{meta}$ utilizes the knowledge on previous tuning tasks, which can greatly accelerate the convergence of the tuning in the target task.
We then use the following ranking loss function
$L$, i.e., the number of misranked pairs, to measure the similarity between previous tasks and the target task:
\begin{equation}
L(M_j,H_T)\!=\!\sum_{j=1}^{n_t}\sum_{k=1}^{n_t}\mathbbm{1}
\!\Big((M_i(\boldsymbol{\theta}_j)\!\leq\! M_i(\boldsymbol{\theta}_k))\! 
\oplus
(y_j\leq y_k)\Big),
\label{e_misrank}
\end{equation}

where $\oplus$ is the exclusive-or operator, $n_t$ denotes the number of tuning tasks  and $M_i(\boldsymbol{\theta}_j)$ means the prediction of $M_i$ on configuration $\boldsymbol{\theta}$.
Based on the ranking loss function, the weight $w_i$ is set to the probability that $M_i$ has the smallest ranking loss on $H_T$, that is, $w_i = \mathbbm{P}(i = \mathop{\arg\min}_jL(M_j,H_T))$.
This probability can be estimated using MCMC sampling~\cite{DBLP:conf/aistats/MartensTY19}.
\begin{table*}[]
\caption{Overview of regression models we evaluated.}\label{tab:appendix-model}
\begin{tabular}{@{}cc@{}}
\toprule
Regression Models                    & Hyper-Parameter                                                                                       \\ \midrule
Random Forest (RF)                   & \textit{n\_estimators, min\_samples\_split, min\_samples\_leaf, max\_features, max\_depth, bootstrap} \\
Gradient Boosting (GB)               & \textit{n\_estimators, min\_samples\_split, min\_samples\_leaf, max\_depth, learning\_rate}           \\
Support Vector Regression (SVR)      & \textit{gamma, C}                                                                                     \\
Nu Support Vector Regression (NuSVR) & \textit{nu, gamma,  C}                                                                                \\
K-Nearest-Neighbours (KNN)           & \textit{n neighbors}                                                                                  \\
Ridge Regression (RR)                & \textit{alpha}                                                                                        \\ \bottomrule
\end{tabular}
\end{table*}

\section{ Construction for database configuration tuning benchmark}

This section presents the procedures to construct the database configuration tuning benchmark and discusses the rationale behind such construction in detail.

\noindent\textbf{Data Collection.}
We collect observation data to train the surrogate model used in the tuning benchmark.
Our ultimate goal is to ensure that optimizers perform similarly using the tuning benchmark as interacting with the real database with workload replay.
In principle, we could construct surrogates using observation data gathered by any means, but of course, we prefer to collect data in a way that leads to the best surrogates. 
Since effective optimizers spend most of their time in high-performance regions of the configuration space, and relative differences between the performance of configurations in such high-performance regions tend to impact which configuration will ultimately be suggested, thus accuracy in this part of the space is more important than in regions of poor performance. Training data should therefore densely sample high-performance regions. 
We thus collect observation data primarily via runs of existing optimizers.
It is also important to accurately identify poorly performing parts of the space to avoid the overly optimistic predictions of performance in poor parts of the space.
We therefore also included performance data gathered by LHS as it can deal effectively with large configuration spaces.
To this end, to collect data for training the surrogate, we used the data gathered by ruining optimizers (e.g., Vanilla BO, DDPG, etc.) tuning the database, and as well as conducting LHS sampling . 


\noindent\textbf{Model Selection and Hyper-parameter Tuning.}
We considered a broad range of commonly used regression algorithms as candidates for our surrogate benchmarks.
Table \ref{tab:appendix-model} details the regression models and their hyper-parameters.
We considered two different tree-based models, Random Forest (RF) and  Gradient  (GB), which have been shown to perform well for non-smooth and high-dimensional problems~\cite{DBLP:conf/lion/HutterHL11}.
We also experimented with k-nearest-neighbors (KNN), ridge regression (RR), and two SVM methods.
We implement all the methods using  scikit-learn~\cite{DBLP:journals/jmlr/PedregosaVGMTGBPWDVPCBPD11} (version 0.22.2) and conduct randomized search on the hyper-parameters.
As shown in Table 9, RF and GB perform similarly. 
As RFs are widely used with
simplicity, we adopt RF as the surrogate for tuning benchmark.
And Table \ref{tab:appendix-hy} presents the hyper-parameters we set for RF.


\begin{table}[]
\caption{Hyper-parameters we use for random forest based surrogate model.}\label{tab:appendix-hy}
\begin{tabular}{@{}ccc@{}}
\toprule
Hyper-Parameter              & SYSBENCH & JOB  \\ \midrule
\textit{n\_estimators}       & 1400     & 800  \\
\textit{min\_samples\_split} & 2        & 2    \\
\textit{min\_samples\_leaf}  & 1        & 1    \\
\textit{max\_features}       & auto     & auto \\
\textit{max\_depth}          & 100      & 100  \\
\textit{bootstrap}           & True     & True \\ \bottomrule
\end{tabular}
\end{table}

\section{More details and results about experiment}
In this section, we present additional experimental details.
\subsection{More Details about Workloads}\label{sec:app-workload}

While we have presented general profile information for workloads in Table ~\ref{wkl} in the paper, we detail the implementation and the reason we select those workloads in this section.

\noindent\textbf{The Reasons for Workload Selection.}
When answering \textbf{Q1} and \textbf{Q2}, we analyze the tuning performances over OLTP and OLAP scenarios.
We  use an OLAP workload -- JOB and an OLTP workload -- SYSBENCH.
The reason is that the two workloads are often adopted in evaluating  database configuration tuning methods and involve the scenarios of an online transaction/analytical processing.
For example, JOB is adopted by QTune~\cite{DBLP:journals/pvldb/LiZLG19} and SYSBENCH is adopted by CDBTune~\cite{DBLP:conf/sigmod/ZhangLZLXCXWCLR19}, QTune~\cite{DBLP:journals/pvldb/LiZLG19} and ResTune~\cite{DBLP:conf/sigmod/ZhangWCJT0Z021}.
When conducting knowledge transfer experiments (\textbf{Q3}), we focus on the OLTP scenarios since there are fewer OLAP workloads suitable for constructing  source workloads of tuning tasks, except JOB and TPC-H.
We choose three OLTP workloads -- SYSBENCH, TPC-C, Twitter as the target tuning workloads, which have been adopted in previous studies.
For example, OtterTune~\cite{DBLP:conf/sigmod/AkenPGZ17}, CDBTune~\cite{DBLP:conf/sigmod/ZhangLZLXCXWCLR19}, and ResTune~\cite{DBLP:conf/sigmod/ZhangWCJT0Z021}) has adopted TPC-C for evaluation and  ResTune has also adopted Twitter.
We use additional four OLTP workloads (i.e., SEATS, Smallbank, TATP, Voter, SIBENCH) as source workloads and configure them with various sizes, read-write ratios as shown in Table ~\ref{wkl}.
SIBench is a microbenchmark designed to explore snapshot isolation in DBMSs~\cite{DBLP:journals/pvldb/JungHFR11}.
Based on our observations, the tuning opportunity for SIBench is limited.
We add SIBench to the source workloads with the purpose of increasing the diversity.

\noindent\textbf{Implementation of The Workloads.}
For JOB, we use the same setup illustrated in \cite{DBLP:journals/pvldb/LeisGMBK015}.
For SYSBENCH, we load 150 tables each of which contains 800000 rows, and adopt the read-write mode.
For workloads from OLTP-Bench, we use the scale factor to determine the data size (e.g., the number of warehouses in TPCC) as shown in Table \ref{tab:appendix-scale}. 
In addition, the parameter \verb|terminal| is set to 64 for each workload. 
We keep other parameters as the default value as OLTP-Bench provided, including \verb|isolation| and \verb|weights| of transactions.

\begin{table}[]
\small
\caption{Scale factors of workload from OLTP-Bench}\label{tab:appendix-scale}
\setlength\tabcolsep{3pt}
\begin{tabular}{cccccccc}
\hline
Workload     & TPCC & Twitter & Smallbank & SIBench & Voter & Seats & TATP \\ \hline
Scale Factor & 200  & 1500    & 10        & 1000    & 10000 & 50    & 100  \\ \hline
\end{tabular}
\end{table}

\subsection{Knob Selection}\label{app-knob}

\begin{table*}[]
\caption{The Top-20 important knobs selected by SHAP for OLTP workloads}\label{tab:appendix-knob}
 \scalebox{0.75}[0.75]{
\begin{tabular}{@{}ccccl@{}}
\toprule
Knob                                     & Type        & Dynamic & Module      & \multicolumn{1}{c}{Description}                                                                  \\ \midrule
innodb\_thread\_concurrency              & Integer     & Yes     & Concurrency & The maximum number of threads permitted inside of InnoDB.                                        \\
innodb\_log\_file\_size                  & Integer     & No      & Logging     & The size in bytes of each log file in a log group.                                               \\
max\_allowed\_packet                     & Integer     & Yes     & Replication & The upper limit on the size of any single message between the MySQL server and clients.          \\
innodb\_io\_capacity\_max                & Integer     & Yes     & IO          & The maximum number of IOPS performed by InnoDB background tasks.                                 \\
tmp\_table\_size                         & Integer     & Yes     & Memory      & The maximum size of internal in-memory temporary tables.                                         \\
query\_prealloc\_size                    & Integer     & Yes     & Memory      & The size in bytes of the persistent buffer used for statement parsing and execution.             \\
max\_heap\_table\_size                   & Integer     & Yes     & Memory      & The maximum size to which user-created memory tables are permitted to grow.                      \\
innodb\_doublewrite                      & Categorical & No      & Memory      & Whether the doublwrite buffer is enabled.                                                        \\
transaction\_alloc\_block\_size          & Interger    & Yes     & Memory      & The amount in bytes by which to increase a per-transaction memory pool which needs memory.       \\
join\_buffer\_size                       & Interger    & Yes     & Memory      & The minimum size of the buffer that is used for joins.                                           \\
innodb\_flush\_log\_at\_trx\_commit      & Categorical & Yes     & Logging     & Controlling the balance between ACID compliance for commit operations and performance.           \\
innodb\_max\_dirty\_pages\_pct\_lwm      & Integer     & Yes     & Logging    & The percentage of dirty pages at which preflushing is enabled to control the dirty page ratio.   \\
innodb\_log\_files\_in\_group            & Integer     & No      & Logging     & The number of log files in the log group.                                                        \\
innodb\_buffer\_pool\_size               & Integer     & Yes     & Memory      & The size in bytes of the buffer pool.                                                            \\
innodb\_online\_alter\_log\_max\_size    & Integer     & Yes     & Logging     & An upper limit on the size of the log files used during online DDL operations for InnoDB tables. \\
key\_cache\_age\_threshold               & Integer     & Yes     & Memory      & The demotion of buffers from the hot sublist of a key cache to the warm sublist.                 \\
binlog\_cache\_size                      & Integer     & Yes     & Memory      & The size of the cache to hold changes to the binary log during a transaction.                    \\
innodb\_purge\_rseg\_truncate\_frequency & Integer     & Yes     & Logging     & The frequency with which the purge system frees rollback segments.                               \\
query\_cache\_limit                      & Integer     & Yes     & Memory      & The minimum size of cached results.                                                              \\
innodb\_sort\_buffer\_size               & Integer     & No      & Memory      & The sort buffer size for online DDL operations that create or rebuild secondary indexes.         \\ \bottomrule
\end{tabular}}
\end{table*}

\begin{table}[t]
\caption{Hardware configurations for more instances.}
\setlength\tabcolsep{10pt}
\small
\label{tab:hardware}
\begin{tabular}{ccccc}
\hline
Instance & A & B & C & D \\ \hline
CPU                & 4 cores    & 8 cores    & 16  cores  & 32  cores  \\
RAM                & 8GB        & 16GB       & 32GB       & 64GB       \\ \hline
\end{tabular}
\end{table}
\noindent\textbf{Top impacting knobs with high tunability for OLTP workloads.} 
We further conduct an experiment using SHAP to generate a ranking of the most impacting knobs across OLTP workloads and hardware instances.
And we use this ranking to conduct an evaluation for \textit{knowledge transfer} component across OLTP workloads in \autoref{sec:exp-transfer} in the paper.
We use the seven OLTP workloads listed in Table ~\ref{wkl} in the paper and perform LHS to collect 1250 samples for each workload on the four hardware instances listed in Table \ref{tab:hardward}.
Then we adopt SHAP to generate an importance ranking respectively and count the number of times that each knob appears in the top 20 of all the rankings to measure their overall importance.
Table~\ref{tab:appendix-knob} shows the top-20 important knobs for OLTP workloads and their brief description.
We believe this ranking could provide database practitioners a guidance for choosing the knobs to tune. 
Dynamic variables can be changed at runtime using the SET statement, while others can only be set at server startup using options on the command line or in an option file.
We have that configuring the maximum number of threads and the size of the log file can contribute to the performance gain the most, which is aligned with the previous analysis ~\cite{DBLP:conf/sigmod/AkenPGZ17,DBLP:conf/sigmod/ZhangWCJT0Z021,DBLP:journals/pvldb/AkenYBFZBP21}.
And 11 of the top-20 knobs are related to memory allocation, which indicates that the default setting of memory allocation in MySQL may not be appropriate across the workloads and hardware instances.
We leave the important knobs ranking for OLAP workloads as future work, as there are fewer OLAP workloads suitable for database tuning tasks, except JOB and TPC-H.
\subsection{Configuration Optimization}

\begin{table*}[t]
\small
\caption{Average ranking of optimizers in terms of the best configuration they found.Bold values are the best.}\label{tab:optimizer-rank}
\setlength\tabcolsep{6.5pt}
\begin{tabular}{cccccccccc}
\hline
Optimizer                                   &                  & Vanilla BO & Oon-Hot BO & Mixed-Kernel BO & SMAC          & TPE  & TURBO & DDPG & GA \\ \hline
\multirow{3}{*}{Small Configuration Space}  & JOB              & 5.00 & 2.67 & \textbf{2.33} & 3.33 & 6.33 & 3.67 & 5.00 & 7.67   \\
                                            & SYSBENCH         & 5.67 & 5.33 & \textbf{2.00} & 3.33 & 5.33 & 4.00 & 5.00 & 5.33   \\
                                            & \textbf{Average} & 5.33 & 4.00 & \textbf{2.17} & 3.33 & 5.83 & 3.83 & 5.00 & 6.50  \\ \hline
\multirow{3}{*}{Medium Configuration Space} & JOB              & 5.33 & 3.33 & 3.00 & \textbf{1.00} & 7.67 & 3.67 & 5.00 & 7.00    \\
                                            & SYSBENCH        & 5.00 & 4.33 & \textbf{1.67} & 1.67 & 6.67 & 4.33 & 6.00 & 6.33     \\
                                            & \textbf{Average} & 5.17 & 3.83 & 2.33 & \textbf{1.33} & 7.17 & 4.00 & 5.50 & 6.67    \\ \hline
\multirow{3}{*}{Large Configuration Space}  & JOB             & 7.00 & 7.00 & 7.00 & \textbf{1.00} & 7.00 & 7.00 & 2.00 & 7.00    \\
                                            & SYSBENCH         & 7.67 & 6.00 & 3.33 & \textbf{1.00} & 6.00 & 3.00 & 4.33 & 4.67    \\
                                            & \textbf{Average} & 7.33 & 6.50 & 5.17 & \textbf{1.00} & 6.50 & 5.00 & 3.17 & 5.83    \\ \hline
\textbf{Overall }                           &                  & 5.94 & 4.78 & 3.22 & \textbf{1.89} & 6.50 & 4.28 & 4.56 & 6.33    \\ \hline
\end{tabular}
\end{table*}

\textbf{Average ranking of optimizers in terms of the best configuration they found.}
While we have presented the average rankings of optimizers in Table 7 in the paper, we detail the rankings on each workload and configuration space as shown in Table \ref{tab:optimizer-rank}.
In addition, we present how we calculate the average ranking of optimizers.
For each workload and configuration space, we run three tuning sessions for an optimizer and sort the three sessions in terms of the best performance they found within 200 iterations.
Then, we rank the optimizers based on the best performance in their best session, and then rank them based on their second session, and lastly the worst session.
Finally, we average the three ranks of an optimizer, which corresponds to a row in Table \ref{tab:optimizer-rank}.

\subsection{Knowledge Transfer}

\begin{figure*}[t]
    \centering
    \includegraphics[width=\textwidth]{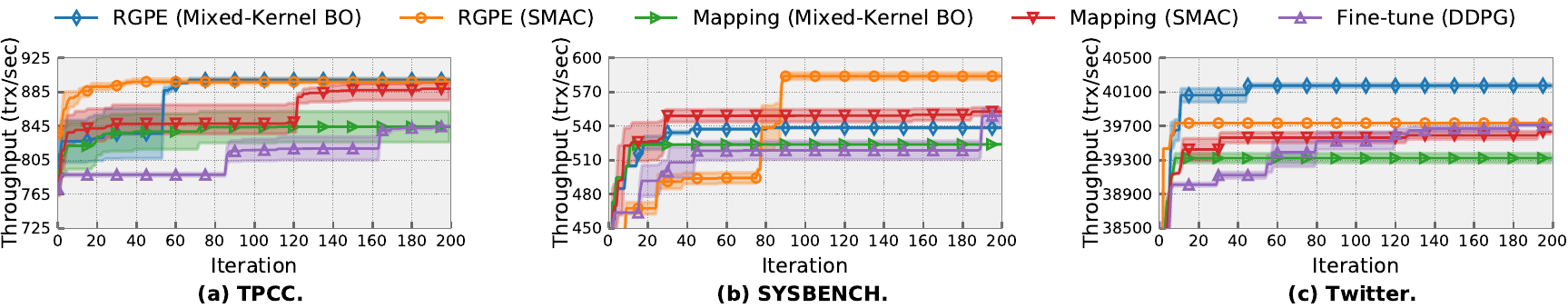}
    \caption{The absolute performance over iteration of each combination of transfer framework and base learner.}
    \label{fig:exp4}
\end{figure*}

We have demonstrated the average performance  enhancement (i.e., PE), speedup, and absolute performance ranking (i.e., APR) in Table 8  and omit the performance plot in the paper due to space constraints.
Figure \ref{fig:exp4} plots the absolute performance over iteration of each baseline (i.e., the combination of transfer framework and base learner).
On TPCC, both RGPE (Mixed-kernel BO) and RGPE (SMAC) find the approximately best performance in 200 iterations, while RGPE (SMAC) has a better speedup.
On SYSBENCH, RGPE (SMAC) finds the best performance, though it takes a few more steps. 
On Twitter, RGPE (Mixed-kernel BO) finds the best performance, and at a fast speed.
In general, we find that the combinations of RGPE and base learners have the best absolute performance as well as speedup.

\section{Evaluations on  PostgreSQL}\label{app:pg}
Different databases have completely different configuration knobs,  including  different  meanings,  types,  names  and value ranges.
In this section, we conduct experiments tuning JOB workload on a different database system, PostgreSQL (v12.7).
We first analyze the important knobs in PostgreSQL and compare the optimizers over small (top-5 important knobs) and medium (top-20 important knob) configuration spaces.
Finally, we present the evaluation results via database tuning benchmark.

\subsection{Knob Selection on PostgreSQL}
Similar to the procedure described in  \autoref{sec:men-knob}, we first collect 3,000 samples for JOB workload on PostgreSQL via Latin Hypercube Sampling and then use SHAP as the importance measurement to select the most important 5 and 20 knobs. 
Before the collection, we find that PostgreSQL is more sensitive to knob settings, i.e., improper knob configurations will lead to  database crashes and terminations of query execution, causing massive failed observations. Therefore, we lower the upper limit of the memory-related knobs based on the hardware settings.

The top-5 important knobs and their brief description are demonstrated in \autoref{tab:pg}, among which two are related to memory, two are related to logging, and one is related to concurrency. 
We observe that the knob with the largest tunability, \textit{max\_worker\_processes} sets the maximum number of background processes that the system can support, which shares a similar function with MySQL's top-1 important knob as shown in Table ~\ref{tab:appendix-knob} 
For \textit{max\_worker\_processes} and \textit{shared\_buffers}, configuring larger values than the default can contribute to the decreasing of latency. 
And this conclusion is on par with the ones on MySQL.
\textit{wal\_buffers} set the amount of shared memory used for WAL data that has not yet been written to disk, with a default setting of -1 that selects a size equal to about 3\% of \textit{shared\_buffers.}, while based on our observation, a higher proportion at about 7\% can achieve the best performance.
To avoid flooding the I/O system with a burst of page writes, \textit{checkpoint\_completion\_target} controls the time of writing dirty buffers during a checkpoint.
With the default value of 0.5 for \textit{checkpoint\_completion\_target}, PostgreSQL is expected to complete each checkpoint in about half the checkpoint interval before the next checkpoint starts. 
In our experiments, a higher value of about 0.8 can improve the performance by reducing the I/O load from checkpoints.
For \textit{bakcend\_flush\_after}, the default is 0, i.e., no forced writeback, while a higher value can limit the amount of dirty data in the kernel's page cache, reducing the likelihood of stalls when a fsync is issued at the end of a checkpoint, or when the OS writes data back in larger batches in the background.


\begin{table*}[]
\caption{The Top-5 important knobs selected by SHAP for JOB on PostgreSQL}\label{tab:pg}
 \scalebox{0.85}[0.85]{
\begin{tabular}{@{}cccl@{}}
\toprule
Knob                   & \multicolumn{1}{c}{Type}  & \multicolumn{1}{c}{Module} & \multicolumn{1}{c}{Description}                          \\ \midrule
max\_worker\_processes & Integer                                    & Concurrency      & Maximum number of background processes that the system can support.           \\
shared\_buffers        & Integer                                    & Memory                     & The number of shared memory buffers used by the server.   \\
wal\_buffers           & Integer                                    & Memory         & The number of disk-page buffers in shared memory for WAL. \\
checkpoint\_completion\_target & Real     & Logging       & Time spent flushing dirty buffers during checkpoint, as fraction of checkpoint interval. \\
backend\_flush\_after          & Integer    & Logging & Number of pages after which previously performed writes are flushed to disk.             \\ \bottomrule
\end{tabular}
}
\end{table*}

\begin{figure*}
  \begin{minipage}[t]{0.5\linewidth}
    \centering
    \includegraphics{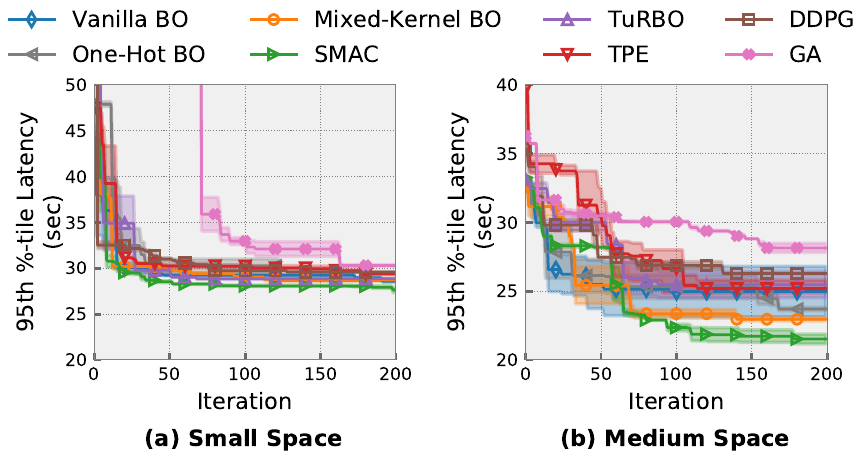}
    \caption{Tuning Performance on PostgreSQL.}
    \label{fig:pg_optimizer}
  \end{minipage}%
  \begin{minipage}[t]{0.5\linewidth}
    \centering
    \includegraphics{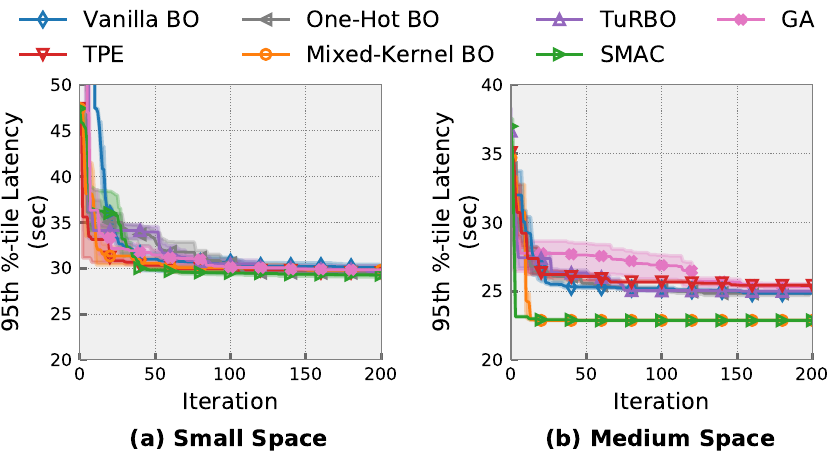}
    \caption{Tuning Performance over surrogate benchmark on PostgreSQL.}
    \label{fig:pg_benchmark}
  \end{minipage}
\end{figure*}

\subsection{Configuration Optimization on PostgreSQL}\label{app-pg}
We further evaluate the optimizers' performance over small and medium spaces for workload JOB on PostgreSQL. 
Figure \ref{fig:pg_optimizer} presents the results, where the conclusion is similar as on MySQL. 
SMAC achieves the best performance in both cases.
For the small space, mixed-kernel BO, one-hot BO, and vanilla BO demonstrate similar convergences since there are no categorical knobs.
For medium space with categorical ones, mixed-kernel BO outperforms one-hot BO  and vanilla BO.
Overall, the BO-based methods outperform RL-based DDPG, while TPE performs poorly on medium space, where there exist knobs that interact with each other (e.g., \textit{checkpoint\_completion\_target} and \textit{checkpoint\_timeout}).

\subsection{Evaluation via Database Tuning Benchmark on  PostgreSQL}\label{app:pg-bench}
To support efficient benchmarking over the configuration space of PostgreSQL, we fit two surrogate models based on RFs over the small and medium configuration space of JOB workload in PostgreSQL. 
And the surrogate model is available online in our repository \footnote{\url{https://github.com/PKU-DAIR/KnobsTuningEA}}.
Figure \ref{fig:pg_benchmark} depicts the best performance found by different optimizers using the tuning benchmark. 
We report means and quartiles across three runs of each optimizer.
We observe that our tuning benchmark yields evaluation results closer to the result in Figure~\ref{fig:pg_optimizer}.
The results on PostgreSQL demonstrate that researchers can conduct algorithm analysis and comparison efficiently under various setups with the help of the tuning benchmark.

\section{Experimental environment and reproduction instructions.}
We conduct all experiments on Aliyun ECS. Each experiment consists of two instances. The first instance is used for the tuning sever, deployed on \verb|ecs.s6-c1m2.xlarge|. The second instance is used for the target DBMS deployment, with four kinds of hardware configurations : \verb|ecs.s6-c1m2.xlarge|, \verb|ecs.s6-c1m2.2xlarge|, \verb|ecs.s6-c1m2.4xlarge|, and \verb|ecs.n4.8xlarge|. The detailed physical memory and CPU information are demonstrated in Table \ref{instance}.
The operation system of each ECS is Linux 4.9. 
The Python version used is 3.7, and the detailed package requirements for our experiments are listed on our GitHub repository. Please check the \verb|requirements.txt| in the root directory.

To reproduce the results of out experiments, please download and install the workloads we use following the instructions in \verb|README.md|, and run \verb|train.py| under directory \verb|script/| with specified arguments like below:

\begin{verbatim}
python train.py --method=VBO --knobs_num=5 --y_variable=lat
--workload=job  --dbname=imdbload --lhs_log=JOB5_VBO.res 
--knobs_config=../experiment/gen_knobs/job_shap.json 
\end{verbatim}

More reproduction details are provided on our online repository\footnotemark[1]. 
\begin{table}[t]
\caption{Hardware configurations for database instances.} \label{instance}
\begin{tabular}{lcc}
\hline
\multicolumn{1}{c}{Type}   & CPU       & RAM  \\ \hline
\verb|ecs.s6-c1m2.xlarge|  & 4 cores   & 8GB  \\
\verb|ecs.s6-c1m2.2xlarge| & 8 cores   & 16GB \\
\verb|ecs.s6-c1m2.4xlarge| & 16  cores & 32GB \\
\verb|ecs.n4.8xlarge|      & 32  cores & 64GB \\ \hline
\end{tabular}
\end{table}

  
\end{document}